# The Impact of Extended $H_2O$ Cross Sections on Temperate Anoxic Planet Atmospheres: Implications for Spectral Characterization of Habitable Worlds

Wynter Broussard[1], Edward W. Schwieterman[1,2], Sukrit Ranjan[2,3], Clara Sousa-Silva[4,5], Alexander Fateev[6], and Christopher T. Reinhard[7]

[1] Department of Earth and Planetary Sciences, University of California, Riverside, CA 92521, USA; abrou009@ucr.edu
[2] Blue Marble Space Institute of Science, Seattle, WA 98154, USA
[3] University of Arizona, Lunar and Planetary Laboratory/Department of Planetary Sciences, Tucson, AZ 85721, USA
[4] Bard College, 30 Campus Rd, Annandale-On-Hudson, NY 12504, USA
[5] Institute of Astrophysics and Space Sciences, Rua das Estrelas 4150-762 Porto, Portugal
[6] Technical University of Denmark, Department of Chemical and Biochemical Engineering, Søltofts Plads 229, Kgs. Lyngby DK 2800, Denmark
[7] School of Earth and Atmospheric Sciences, Georgia Institute of Technology, Atlanta, GA 30332, USA

## Abstract

JWST has created a new era of terrestrial exoplanet atmospheric characterization, and with it, the possibility to detect potential biosignature gases like $CH_4$. Our interpretation of exoplanet atmospheric spectra, and the veracity of these interpretations, will be limited by our understanding of atmospheric processes and the accuracy of input modeling data. Molecular cross sections are essential inputs to these models. The photochemistry of temperate planets depends on photolysis reactions whose rates are governed by the dissociation cross sections of key molecules. $H_2O$ is one such molecule; the photolysis of $H_2O$ produces OH, a highly reactive and efficient sink for atmospheric trace gases. We investigate the photochemical effects of improved $H_2O$ cross sections on anoxic terrestrial planets as a function of host star spectral type and $CH_4$ surface flux. Our results show that updated $H_2O$ cross sections, extended to wavelengths >200 nm, substantially impact the predicted abundances of trace gases destroyed by OH. The differences for anoxic terrestrial planets orbiting Sun-like host stars are greatest, showing changes of up to 3 orders of magnitude in surface CO levels, and over an order of magnitude in surface $CH_4$ levels. These differences lead to observable changes in simulated planetary spectra, especially important in the context of future direct-imaging missions. In contrast, the atmospheres of planets orbiting M-dwarf stars are substantially less affected. Our results demonstrate a pressing need for refined dissociation cross-section data for $H_2O$, where uncertainties remain, and other key molecules, especially at mid-UV wavelengths >200 nm.

## 1. Introduction

Recent surveys and statistical analyses have revealed that rocky terrestrial planets are common in the galaxy (Dressing & Charbonneau 2015; Bryson et al. 2020; Dattilo et al. 2023). The launch of JWST, upcoming ground-based instruments and facilities, and the recommendations of the 2020 Astronomy & Astrophysics Decadal Survey have invigorated prospects for the detailed atmospheric characterization of rocky planets in the coming years and decades. Rocky exoplanets with secondary atmospheres, including Earth, our only known example of a life-bearing world, are central to the search for life outside our solar system (Kaltenegger 2017; Schwieterman et al. 2018).

Photochemical studies are needed to understand both the spectroscopic observables and the environments from which these observables originate on temperate, Earth-like planets. For these photochemical studies to be robust, photochemical models require a wide range of inputs, including chemical reaction rates, molecular cross sections, stellar spectra, dry and wet deposition rates, and mixing parameterizations. Models that have been calibrated for a narrow set of conditions, most often the modern Earth or solar system worlds, can err in their predictions when inputs are not sufficiently complete for vastly different boundary conditions. Moreover, models that have not harmonized their photochemical inputs will yield incompatible predictions. From an observational standpoint, interpretations or retrievals using divergent model inputs will produce conflicting results with reported uncertainties that include unacknowledged systematic errors (e.g., Niraula et al. 2022).

Recently, Ranjan et al. (2020) showed that updates to $H_2O$ cross sections can meaningfully impact predictions of trace gas chemistry on anoxic, prebiotic planets. Specifically, past prescriptions for photochemical models have truncated the $H_2O$ photolysis cross sections at ∼200–208 nm, where the $H_2O$ opacity falls below the typical scattering opacity of the atmosphere, while the ground-state quantum limit for the H–OH bond at room temperature is ∼240 nm. Historically, this truncation has occurred since $H_2O$ is only marginally spectrally active in this region, and its opacity is far too weak to be used to detect water. Despite this, the 200–240 nm range is consequential for atmospheric chemistry in thick, anoxic atmospheres because stellar mid-UV (MUV; 200–300 nm) photons can penetrate into the $H_2O$-rich troposphere, while higher-energy far-UV (FUV; <200 nm) photons are shielded by overlying $CO_2$ in the $H_2O$-poor stratosphere. (In $O_2$-rich atmospheres like that of modern Earth, most of these photons are shielded in the upper atmosphere by overlying $O_2$ and $O_3$ and are therefore of much less importance for $H_2O$ photolysis.)

$H_2O$ photolysis, following Equation (1), is a particularly important reaction because it serves as the main source of OH in these anoxic atmospheres (in contrast, in $O_2$-rich atmospheres, OH is primarily sourced from the reaction $O(^1D)$ + $H_2O \rightarrow 2OH$, where the $O(^1D)$ is derived from the photolysis of tropospheric ozone). Ranjan et al. (2020) used newly measured $H_2O$ cross sections to show that these previously neglected MUV photons liberate OH radicals in the troposphere, which subsequently react with trace gases such as $CH_4$ and assist in catalyzing the recombination of CO and $O_2$ back into $CO_2$ (via the reaction CO + OH → $CO_2$ + H). Consequently, these updates to the $H_2O$ cross sections harmonized divergent photochemical predictions for the oxidation states of prebiotic anoxic terrestrial planet atmospheres and yielded insights into the trace gas abundance and delivery of reducing compounds to the surface of such worlds. These types of advances have been highlighted as a priority for decades (Wen et al. 1989), and are essential to the future characterization of temperate terrestrial exoplanets with secondary atmospheres.

$$H_2O + h\nu \rightarrow H + OH. \quad (1)$$

$CH_4$ is a key biosignature gas that JWST could plausibly detect in the atmospheres of rocky planets orbiting cool M-dwarf stars (Krissansen-Totton et al. 2018; Thompson et al. 2022). The thermodynamic and kinetic disequilibria between atmospheric $CH_4$ and other atmospheric constituents underpins its potential value as a biosignature (Sagan et al. 1993; Wogan & Catling 2020; Krissansen-Totton et al. 2022; Thompson et al. 2022). However, in order to accurately interpret retrieved $CH_4$ abundances, we must understand the relationship between $CH_4$ surface flux and $CH_4$ surface mixing ratio. In other words, for a given planet-star-atmosphere configuration, a measured mixing ratio corresponds to a $CH_4$-production rate, and that production rate offers insights into plausible $CH_4$ sources. Because we know values for various biogenic and abiogenic $CH_4$ fluxes on Earth, it is thus possible in principle to assess the likelihood of biogenic production (Thompson et al. 2022). However, determining the required rate of photochemical production (which is equivalent to the rate of photochemical destruction in steady state), requires accurate photochemical inputs. In addition, $H_2O$ inputs will impact not only $CH_4$, but also other important trace species, including CO and $O_2$, which may likewise be critical for interpreting potential biosignature gas combinations (Harman et al. 2015; Schwieterman et al. 2019; Ranjan et al. 2023).

It is not our intent here to fully evaluate the plausibility of assessing abiotic versus biotic $CH_4$ in a habitable exoplanet, but rather to illustrate the large discrepancies that are possible with incomplete fundamental photochemical inputs and how this may affect the interpretation of future exoplanetary observations. The updated $H_2O$ cross sections from Ranjan et al. (2020) improved upon previously nonexistent measurements for $H_2O$ at habitable temperatures in the MUV, but they still serve as a more conservative estimate, leading to smaller $H_2O$ photolysis rates than may be physically realistic. Additionally, these cross sections have yet to be independently confirmed or reproduced by other groups. By evaluating the impacts of the updated $H_2O$ cross sections and demonstrating the sensitivity of the photochemical model to this vital input, we aim to emphasize the importance of using extended UV cross sections in models of anoxic terrestrial planets and highlight the need to take additional measurements of $H_2O$ cross sections in the MUV. More broadly, we hope to show that predictions of atmospheric photochemistry on temperate terrestrial planets can depend sensitively on the long-wavelength tail of the photodissociation opacities of major species, which are often prematurely truncated or extrapolated ad hoc rather than measured, demonstrating critical urgency for improved laboratory and ab initio investigations of these fundamental photochemical inputs (e.g., Fortney et al. 2016, 2019).

In this paper, we conduct a series of $CH_4$ flux-abundance sensitivity tests using various $H_2O$ cross-section prescriptions. We examine both the effects of these cross sections on $CH_4$ and on associated photochemical products CO and $O_2$. We perform these sensitivity tests for planets orbiting FGKM-type stars to determine how the $CH_4$ flux trace gas abundance relationship varies with stellar type for anoxic, temperate, Earth-like exoplanet atmospheres. In Section 2, we describe the methods used to conduct these sensitivity tests, and describe the photochemical model and planetary atmosphere scenarios used to test the sensitivity of these inputs under a range of conditions. In Section 3, we report our results, including impacts to trace gas species and simulated emission, reflection, and transmission spectra. We discuss the implications of our results in Section 4 and conclude in Section 5.

## 2. Methods

### 2.1. Cross-section Prescriptions

We conduct a $H_2O$ cross-section sensitivity test using three different prescriptions, shown in Figure 1. This includes the *old* cross sections from Kasting & Walker (1981), which were the default in the Atmos photochemical code prior to the updated measurements from Ranjan et al. (2020). The *new* cross sections are the extrapolation prescription from Ranjan et al. (2020). From 215–230 nm, the measured $H_2O$ cross sections from Ranjan et al. (2020) showed a deviation from the expected log-linear decrease with wavelength. This deviation might have come from the data approaching the noise limits of the instrumentation; to reduce the chance that their predicted water absorption was an overestimate, they modified their measured cross sections into two different prescriptions: a cutoff prescription and an extrapolated prescription. The cutoff prescription took the measured cross sections, but truncated them beyond about 216 nm (Ranjan et al. 2020). The extrapolated cross sections replaced the measured cross sections beyond 205 nm with a log-linear extrapolation of the data from 186–205 nm, to account for the noisiness of the data beyond 205 nm; thus, the new cross sections, and their photochemical impact, should be considered conservative estimates (we evaluate the cutoff versus the extrapolated cross-section prescriptions in Appendix C). The "abrv" prescription is an abbreviated version of the new cross sections, terminating at 200 nm; this termination wavelength was the recommendation for $H_2O$ (Sander et al. 2011) prior to Ranjan et al. (2020), though we note that our $H_2O$ cross sections are not identical to those of Sander et al. (2011) at wavelengths $<200$ nm. Using the new cross sections with a 200 nm cutoff allows us to conduct tests *all else being equal,* while minimizing the total number of possible permutations.

For the purpose of these sensitivity tests, we use the $CO_2$ cross sections from Kasting & Walker (1981) rather than the more recent cross sections from Lincowski et al. (2018), which are extended ad hoc with a log-extrapolation to 225 nm. This choice allows our results to be consistent with the choice of $CO_2$ cross sections from Ranjan et al. (2020). We have, however, examined our results under the assumption of the Lincowski et al. (2018) cross sections and find the major conclusions presented here are unaffected.

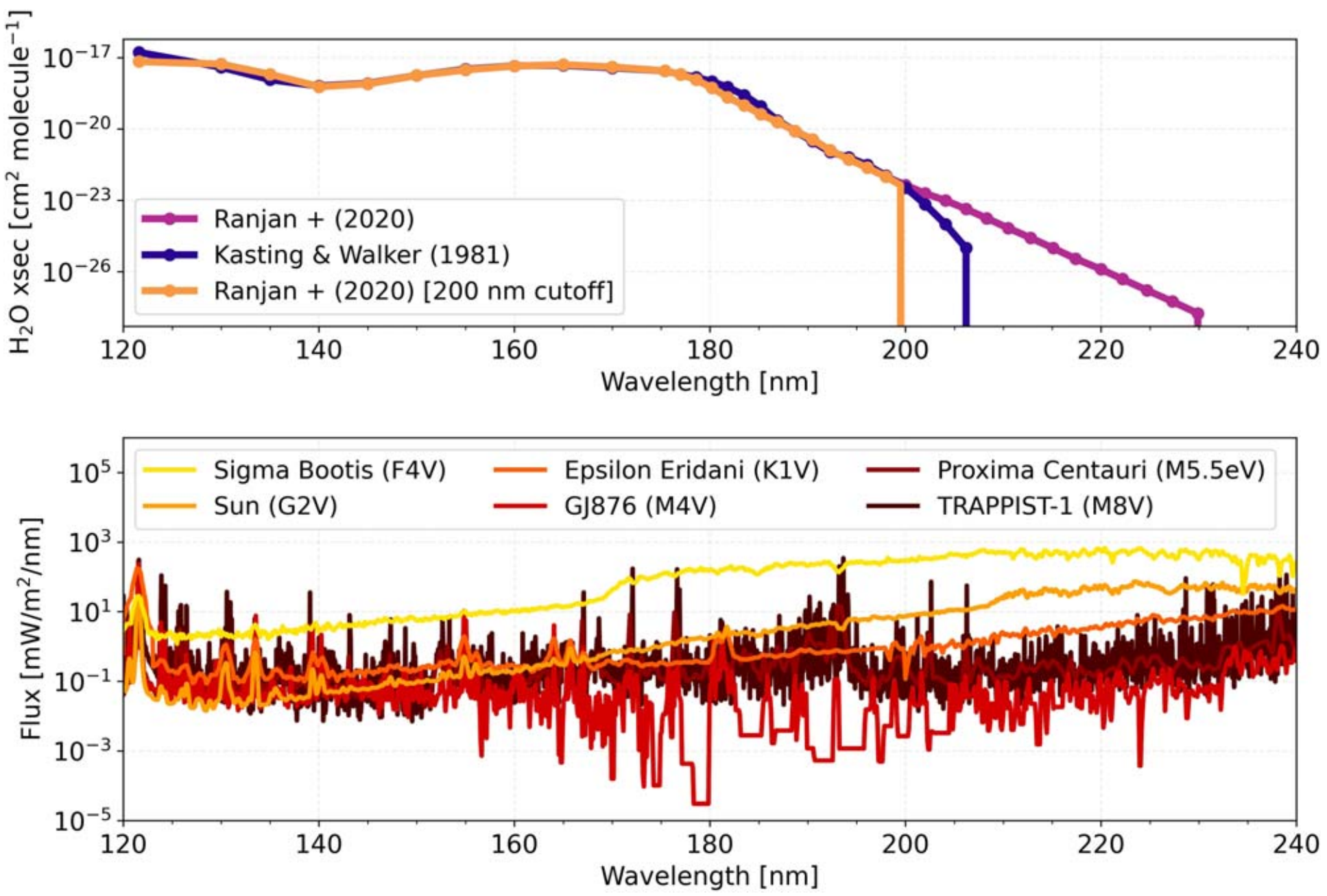

**Figure 1.** (Top) $H_2O$ cross-section prescriptions from 120–240 nm. (Bottom) Spectral energy distribution of the stars modeled in this study, from 120–240 nm.

### 2.2. Photochemical Model Description

The photochemical model used in this research is Atmos, a one-dimensional coupled photochemical-climate model (though for the purpose of this research, only the photochemical portion of the model was used). First developed by Kasting et al. (1979), Atmos is frequently updated to keep the chemical reactions rates current (Burkholder et al. 2019), and to improve its capabilities (Zahnle et al. 2006; Arney et al. 2016; Lincowski et al. 2018). Atmos is often used to model terrestrial exoplanet atmospheres (Arney et al. 2016, 2018; Felton et al. 2022; Schwieterman et al. 2022).

Atmos takes in prescribed planetary and stellar parameters, as well as gas species boundary conditions and various fundamental inputs. The atmosphere is divided into 200 equally spaced vertical layers of 0.5 km; flux and mass continuity equations are simultaneously solved at each layer using the reverse Euler method. Once converged, Atmos returns volume mixing ratios for relevant gases at each atmospheric layer.

For the purpose of this research, we used the Archean + haze template described in Arney et al. (2016), with the atmospheric boundary conditions found in Table 2 in Appendix A. We used a modified temperature-pressure profile, setting the surface pressure at 1 bar, and the surface temperature at either 275 or 300 K. The tropospheric water vapor profile is calculated assuming a surface relative humidity of 70% (Manabe & Wetherald 1967). Two surface temperatures were chosen to illustrate the relative impact of the differing $H_2O$ cross sections for cool, more water-poor tropospheres versus warmer, more water-rich tropospheres (within the habitable range). The temperature decreases linearly from the surface to the tropopause (set at 11 km), at which point it becomes an isothermal profile, fixed at 180 K.

### 2.3. Spectral Model Description

To simulate the spectra of the modeled atmospheres, we use the Spectral Mapping Atmospheric Radiative Transfer code (SMART; Meadows & Crisp 1996; Crisp 1997). SMART takes the temperature-pressure profile and altitude-dependent gas volume mixing ratios from Atmos as input. Contained within SMART is the Line-by-Line Absorption Coefficients (LBLABC) model, which uses spectral line parameters from the HITRAN molecular spectroscopic database (Gordon et al. 2022), but can also incorporate user-supplied data when HITRAN line lists are unavailable. LBLABC calculates the wavelength-dependent opacities from each gas and returns these to SMART. SMART then goes through radiative transfer calculations at each atmospheric layer, incorporating the wavelength-dependent opacities as well as individual gas cross sections and collision-induced absorption coefficients. SMART has been verified for several solar system objects (Tinetti et al. 2006; Robinson et al. 2011; Arney et al. 2014) and is frequently used for simulating exoplanet spectra (Charnay et al. 2015; Lincowski et al. 2018; Lustig-Yaeger et al. 2019; Meadows et al. 2023).

### 2.4. Stellar Spectra

Photochemical reactions, driven by the actinic flux of the host star, are dependent on the host star's spectrum. To model a broad range of stellar types, we chose six representative stars; three Sun-like (FGK) stars, and three M-type stars. The Sun-like stars include $\sigma$ Bootis (Segura et al. 2003), an F-type star, the Sun, a G-type star (Thuillier et al. 2004), and $\epsilon$ Eridani (Segura et al. 2003), a K-type star. The three M-type stars include GJ 876 (M4V) (France et al. 2016; Loyd et al. 2016, "v22"; Youngblood et al. 2016), Proxima Centauri (M5.5eV) (Shkolnik & Barman 2014; Loyd et al. 2018; Peacock et al. 2020), and TRAPPIST-1 (M8V) (Peacock et al. 2019a, 2019b). $H_2O$ photolysis is driven by the UV flux of the host star. Table 1 provides the stellar properties of the six host stars modeled here, and Figure 1 shows their UV spectral energy distributions.

**Table 1**
Stellar Properties

| Star | Spectral Type | $T_{\mathrm{eff}}$ (K) | Luminosity ($L_\odot$) | Stellar Radius ($R_\odot$) | Distance (pc) |
|---|---|---|---|---|---|
| $\sigma$ Bootis | F4V | 6435 | 3.1541 | 1.4307 | 15.8 |
| Sun | G2V | 5780 | 1 | 1 | ⋯ |
| $\epsilon$ Eridani | K1V | 5039 | 0.32 | 0.735 | 3.2 |
| GJ 876 | M4V | 3129 | 0.0122 | 0.3761 | 4.69 |
| Proxima Centauri | M5.5 eV | 2992 | 0.001567 | 0.147 | 1.3 |
| TRAPPIST-1 | M8V | 2559 | 0.000524 | 0.117 | 12.1 |

### *2.5. Planetary Scenario*

The atmospheres modeled in this work are $N_2$–$CO_2$-$H_2O$ atmospheres, consistent with those typically assumed for habitable worlds (e.g., Kopparapu et al. 2013) and within the plausible bounds of the atmospheric composition of the Archean Earth (Arney et al. 2016; Catling & Zahnle 2020). Boundary conditions for each species, including surface flux, volume mixing ratio, and dry deposition rate (if prescribed) are given in Table 2. A key finding from Ranjan et al. (2020) was that their results were sensitive to the assumption of global redox balance. On an anoxic, uninhabited planet, the net delivery of oxidants or reductants to the ocean should be met by a return flux to the atmosphere, assuming no abiotic mechanism to remove excess reduced (or oxidized) materials from the system (e.g., through burial) (Harman et al. 2015; Ranjan et al. 2020). On a planet with life, the production and burial of organic carbon can lead to a net removal of reductants from the ocean-atmosphere system, and the principle of global redox balance does not necessarily apply. Because our focus here is modeling a biotic scenario, we do not enforce global redox balance. We present a second set of simulations assuming abiotic boundary conditions in Appendix B, which also do not enforce global redox balance, in part to be directly comparable with the biotic simulations presented in the main text. However, we tested the sensitivity to imposing global redox for select scenarios and found our results for surface $CH_4$ mixing ratios to be sensitive to the assumption of global redox balance, similar to the conclusions from Ranjan et al. (2020; see Appendix B for additional details).

Beyond the boundary conditions found in Table 2, we use $N_2$ as a filler gas, and the total surface pressure is set to 1 bar. We adopt a $CO_2$ mixing ratio of 2%. Figure 2 shows an example profile plot from our $H_2O$ cross-section sensitivity tests, showing resulting altitude-dependent mixing ratios of key gases using the 275 K new and abbreviated cross sections for the Sun, with the $CH_4$ flux = $1.1 \times 10^{10}$ molecules cm$^{-2}$ s$^{-1}$, which is about one-tenth of Earth's modern $CH_4$ flux.

To conduct the $CH_4$ sensitivity tests we vary the $CH_4$ surface flux, starting at a surface flux of $1.0 \times 10^9$ molecules cm$^{-2}$ s$^{-1}$ ($\sim 2.67 \times 10^{-1}$ Tmol yr$^{-1}$), which is around the upper limit of $CH_4$ flux for serpentinizing systems (Thompson et al. 2022), to a surface flux of $1.0 \times 10^{11}$ molecules cm$^{-2}$ s$^{-1}$ ($\sim 2.67 \times 10^1$ Tmol yr$^{-1}$), representing a roughly Earth-like $CH_4$ flux. (Earth's current methane production levels are around 30 Tmol yr$^{-1}$, or $\sim 1.12 \times 10^{11}$ molecules cm$^{-2}$ s$^{-1}$ (Thompson et al. 2022)).

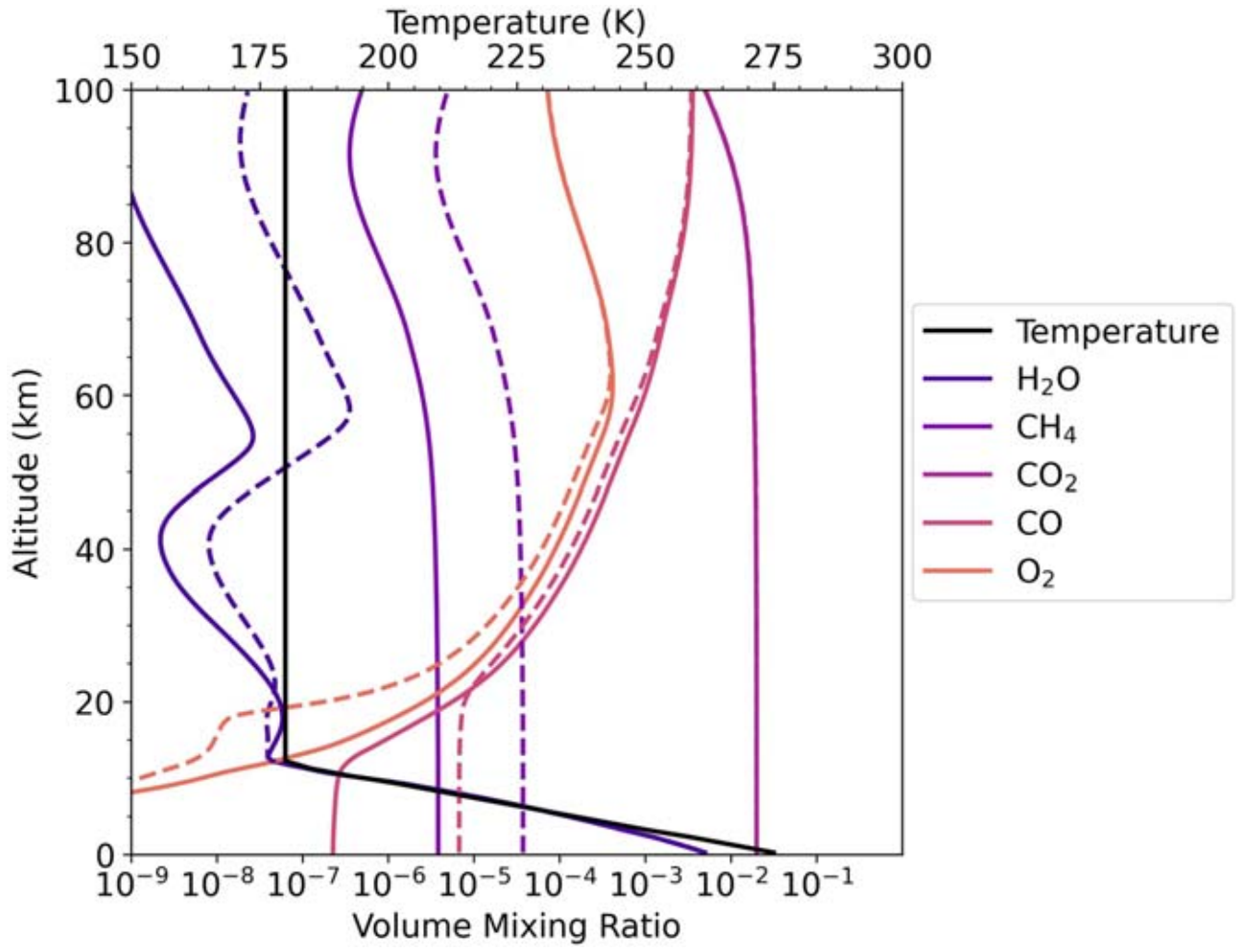

**Figure 2.** Example profile plot for an anoxic habitable planet orbiting the Sun, $CH_4$ = $1.1 \times 10^{10}$ molecules cm$^{-2}$ s$^{-1}$. Solid lines show the altitude-dependent volume mixing ratios of key gases modeled using the new $H_2O$ cross sections; dashed lines show the altitude-dependent volume mixing ratios of key gases modeled using the abbreviated $H_2O$ cross sections.

## 3. Results

### *3.1. $H_2O$ Cross–section Impacts on Trace Gas Relationships*

Figure 3 shows the sensitivity of the $CH_4$ surface volume mixing ratio to $CH_4$ flux, with the various $H_2O$ cross-section prescriptions and host star stellar types. Figure 4 shows these results for the surface CO volume mixing ratios, and Figure 5 the results for the surface $O_2$ mixing ratios. Figures 6, 7, and 8 more directly compare the trace gas surface mixing ratios at specific $CH_4$ fluxes of $10^9$, $10^{10}$, and $10^{11}$ molecules cm$^{-2}$ s$^{-1}$. Generally, the largest differences between atmospheric trace gas volume mixing ratios modeled using the various $H_2O$ cross sections occur in the troposphere, closest to the surface, where the atmosphere is most dense. For this reason, we describe the impacts on the surface volume mixing ratio of $CH_4$, CO, and $O_2$. Appendix D describes the impacts to trace gas column densities; notably, Figure 18 shows that $CH_4$ column density tracks the same relationship as the $CH_4$ volume mixing ratio, thus conclusions drawn from Figure 3 will hold in either case.

For surface $CH_4$ volume mixing ratios, the different host star scenarios show differences of up to an order of magnitude, with the largest differences occurring between the new and abbreviated cross sections around a $CH_4$ flux of $10^{10}$ molecules cm$^{-2}$ s$^{-1}$, for the G-type host star. Near the higher $CH_4$ fluxes, the K- and M-type host stars show surface $CH_4$ mixing ratios that approach a common value independent of the cross-section prescription. This is consistent with the OH sink becoming saturated. This is true regardless of the cross-section prescription since the MUV flux of these host stars is so low that there are not enough MUV photons available to produce OH through $H_2O$ photolysis and the amount of OH produced is not able to keep up with the increasing amount of $CH_4$ in the atmosphere. Thus, the OH sink becomes overwhelmed, allowing trace gases like $CH_4$ to build up more easily. For the K- and M-type host stars, smaller MUV fluxes lead to the availability of MUV photons being the limiting factor in the production of OH through $H_2O$ photolysis. As more $CH_4$ is introduced into the atmosphere, OH is more effectively consumed, allowing trace gases to build up that would otherwise be *scrubbed out*. With the greater MUV flux of the G-type host star, OH production reflects the balance between MUV photon availability and $H_2O$ availability; with the drier troposphere of the 275 K regime, we can see the OH sink does not become saturated except at the highest $CH_4$ fluxes. In the 300 K regime there is more $H_2O$ available for photolysis, with the result that OH production is able to keep up with the increasing $CH_4$ flux, preventing saturation of the OH sink. In the case of the F-type host star, the MUV flux fuels enough $H_2O$ photolysis that OH production can keep up with the increasing $CH_4$ flux. We then see much less surface $CH_4$ buildup (note the more limited scaling of the *y*-axis for the F-type host star in Figure 3). The difference between the $H_2O$ cross-section cases is more apparent for the drier 275 K temperature regime, and differences increase for larger $CH_4$ fluxes. Notably, there is some apparent jaggedness to the behavior of the M-type host stars around the middle of the $CH_4$ flux parameter space for the 275 K temperature regime. This behavior is due to the threshold by which the $CH_4$ collapses the $O_2$ levels, and is very sensitive to the choice of input parameters, including the $CH_4$ flux and cross-section prescription.

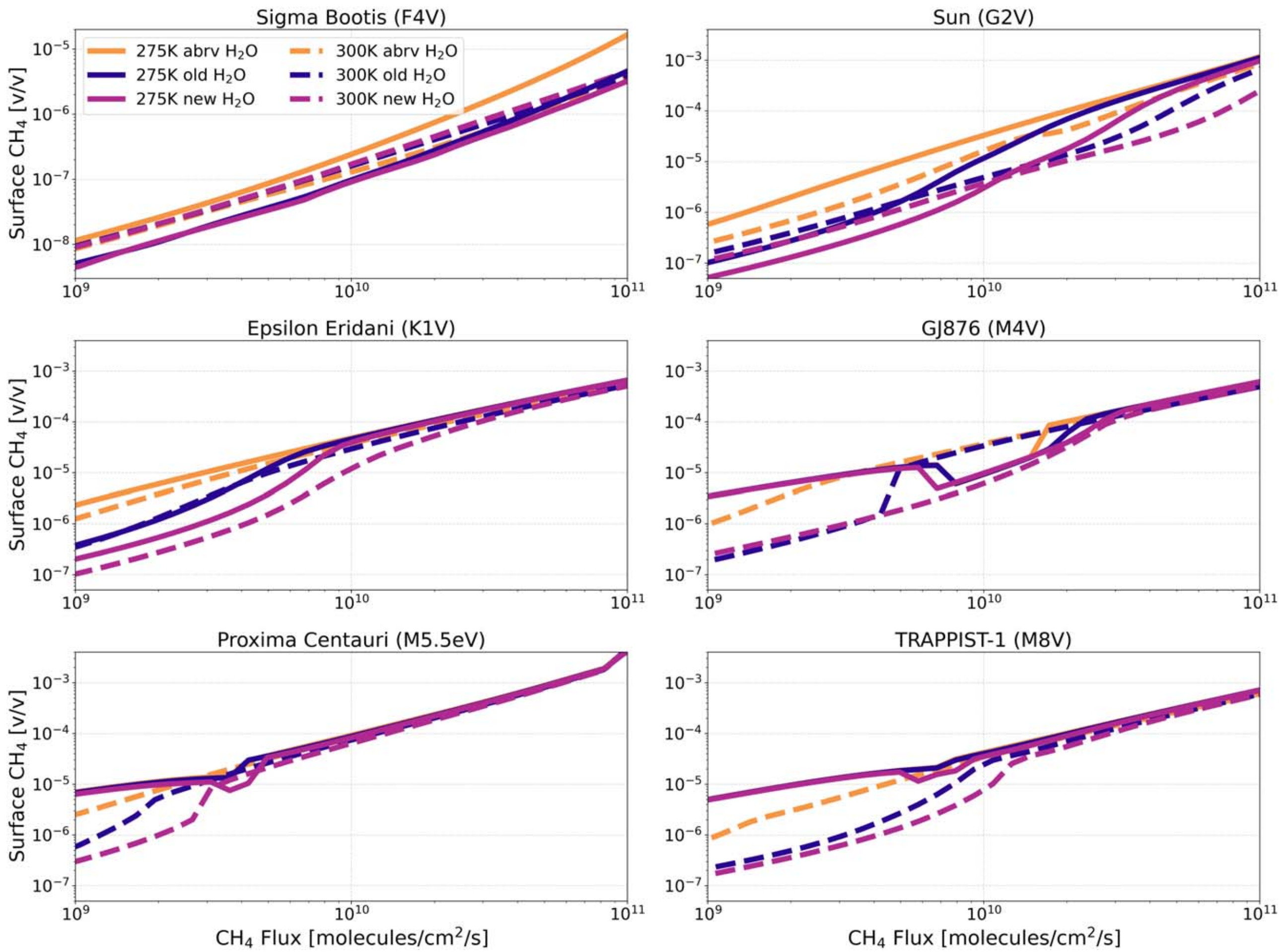

**Figure 3.** $H_2O$ cross-section sensitivity test; surface $CH_4$ vs. $CH_4$ flux for anoxic habitable planets orbiting FGKM-type host stars.

Surface CO volume mixing ratios (Figure 4) can show differences of up to 3 orders of magnitude (as in the case of the F-type host star). CO is a photochemical product of $CH_4$ processing (Schwieterman et al. 2019), so it holds that the maximum differences occur at the lowest $CH_4$ fluxes. As with surface $CH_4$, the surface CO mixing ratios for the K- and M-type host stars approach a common value independent of the choice of temperature regime or cross section with increasing $CH_4$ flux, or in the case of the 275 K regime for the G-type host star. Here, the F-type host star shows the greatest sensitivity to $H_2O$ cross-section prescription.

In Figure 5, we see large variations in the surface $O_2$, with up to 5 orders of magnitude difference between the abbreviated and new cross sections in the 300 K regime for the F-type host star, and around 3 orders of magnitude difference for the G-type host star. However, it is important to note that these high order of magnitude variations are occurring at very small $O_2$ mixing ratios, e.g., $7.9 \times 10^{-15}$ versus $4.9 \times 10^{-10}$ (as for the F-type host star with a $CH_4$ flux of $10^9$ molecules $cm^{-2}$ $s^{-1}$, shown in Figure 8). Both these cases are very anoxic, and neither would result in observable spectral features from $O_2$ or $O_3$. For the M-type host stars, the surface $O_2$ volume mixing ratios appear to be more strongly temperature-dependent than the other stellar types, with over 5 orders of magnitude in variation between the 275 and 300 K surface temperatures and associated tropospheric water vapor, but very little change between the old, new, and abbreviated $H_2O$ cross-section prescriptions. This is primarily attributable to much less $H_2O$ photolysis at the much lower MUV flux of the M-dwarfs relative to the higher-mass stars, which renders changes to the MUV end of the $H_2O$ cross sections less noticeable.

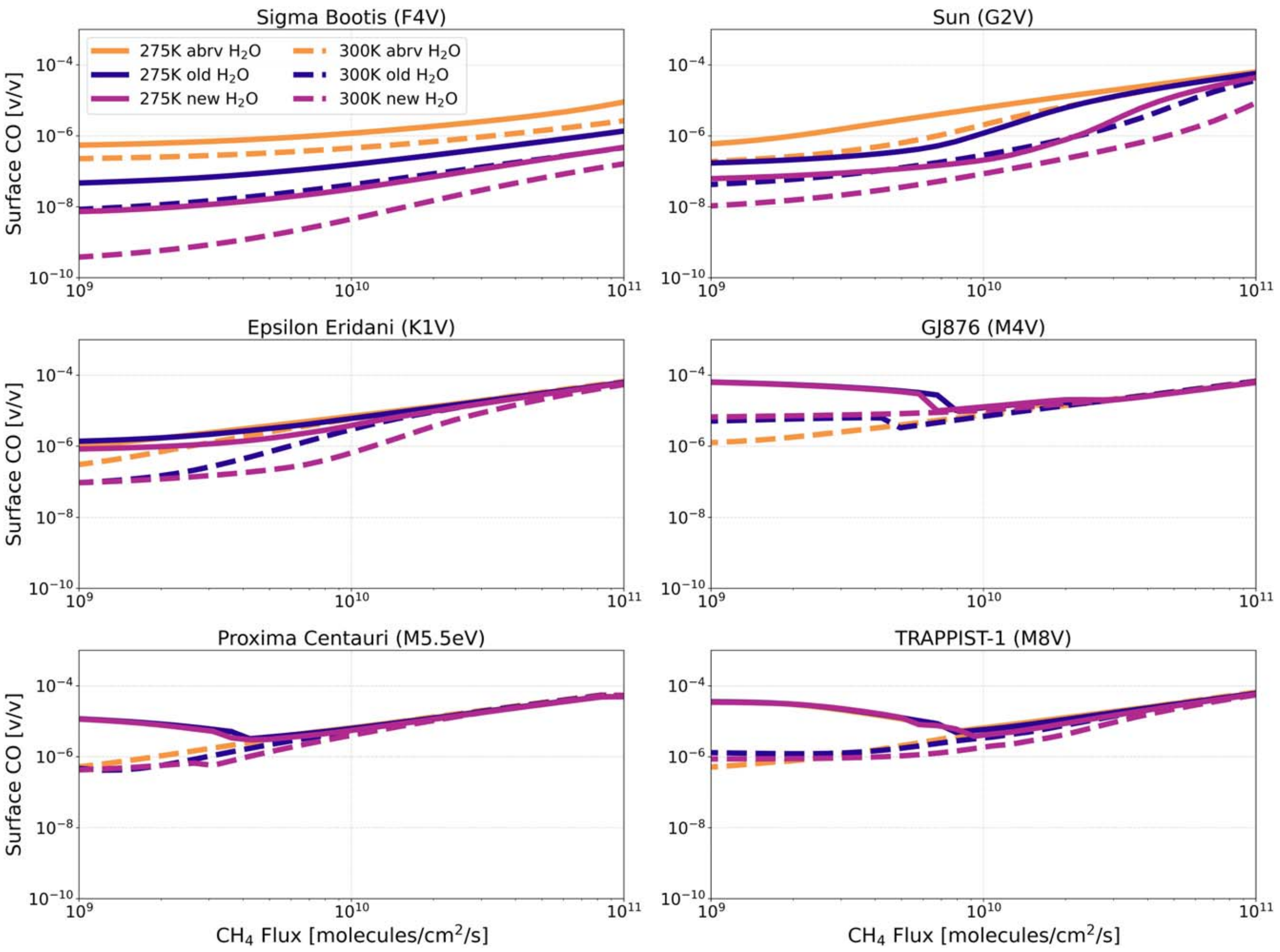

**Figure 4.** $H_2O$ cross-section sensitivity test; surface CO vs. $CH_4$ flux for anoxic habitable planets orbiting FGKM-type host stars.

The behavior of the surface $O_2$ is somewhat different from the other trace gases. Since $O_2$ is indirectly consumed by $CH_4$ through oxidation reactions, increasing the $CH_4$ flux in this scenario leads to decreasing surface $O_2$. This interaction is clearest with the M-type host stars, where the consumption of $O_2$ through reactions with $CH_4$ is strongly temperature dependent. In the 300 K regime, there is enough OH being produced from $H_2O$ photolysis that differences between the cross sections are more apparent. In the 275 K regime, the biggest difference resulting from the various cross-section prescriptions is the flux level where $CH_4$ starts to lead to a stepwise decrease in $O_2$; with the new cross sections, the $O_2$ drops off at slightly smaller $CH_4$ fluxes, e.g., at $5 \times 10^9$ molecules $cm^{-2}$ $s^{-1}$ for the new cross sections, as opposed to $7 \times 10^9$ molecules $cm^{-2}$ $s^{-1}$ for the abbreviated cross sections, in the case of TRAPPIST-1 as the host star.

Figure 9 shows relevant chemical reaction rates for the same simulations shown in Figure 2. The difference between the new and abbreviated $H_2O$ photolysis rates is minimal near where $CO_2$ photolysis is maximal, and greatest where most of the water is present near the surface. The increased $H_2O$ photolysis in the troposphere for the new cross sections corresponds with greater production of OH, which then leads to the greater consumption of CO, $CH_4$, $CH_3$, and other trace gases via their reactions with OH. $H_2O$ photolysis in the upper atmosphere is greater for the abbreviated cross sections than for the new cross sections since as we can see in Figure 2, there is more $H_2O$ in this part of the atmosphere (hence the generally larger OH reaction rates for the abbreviated cross sections in the upper atmosphere). Note that stratospheric $CH_4$ leads to the production of stratospheric $H_2O$ (Segura et al. 2005).

### *3.2. Spectral Sensitivity*

Depending on the stellar type of the planet's host star, the varying atmospheric trace gas abundances that result from the choice of $H_2O$ cross-section prescription will lead to observable differences in the resulting emission, reflection, and transmission spectra. To show an example of this, we have simulated the emission, reflection, and transmission spectra for a planet orbiting the Sun with a $CH_4$ flux of $1.1 \times 10^{10}$ molecules $cm^{-2}$ $s^{-1}$, equivalent to roughly 3 Tmol $year^{-1}$. This flux is near the maximum possible non-biogenic $CH_4$ flux given by Krissansen-Totton et al. (2018) and is well within plausible bounds for a methanogenic biosphere. This flux thus lies near a critical threshold for differentiating an abiotic world from an inhabited one. These spectra were generated assuming a cloud coverage of 50% clear sky, 25% cirrus, and 25% alto-stratus.

In the emission spectrum shown in Figure 10, we have labeled features from $CO_2$ and $CH_4$. Around 7 $\mu$m, we can see that the increased $CH_4$ resulting from the abbreviated cross sections results in a deeper $CH_4$ feature. In the reflection and transmission spectra (Figures 11 and 12), we have labeled spectral features from $CH_4$, $CO_2$, CO, and $H_2O$. Most of the deeper features of the abbreviated reflection spectrum are due to greater $CH_4$, though there is a small contribution from CO at around 2.35 $\mu$m. These results are particularly relevant for the upcoming Habitable Worlds Observatory (HWO). In the transmission spectra, most of the features are again due to differences in $CH_4$ abundance, though there is a very small feature around 4.6 $\mu$m due to CO. Although transmission spectra of rocky exoplanets around Sun-like stars are not attainable with JWST, and the proposed HWO will primarily characterize planets in reflected light, it may be possible to characterize some targets through transmission spectroscopy depending on the final HWO design (The LUVOIR Team 2019).

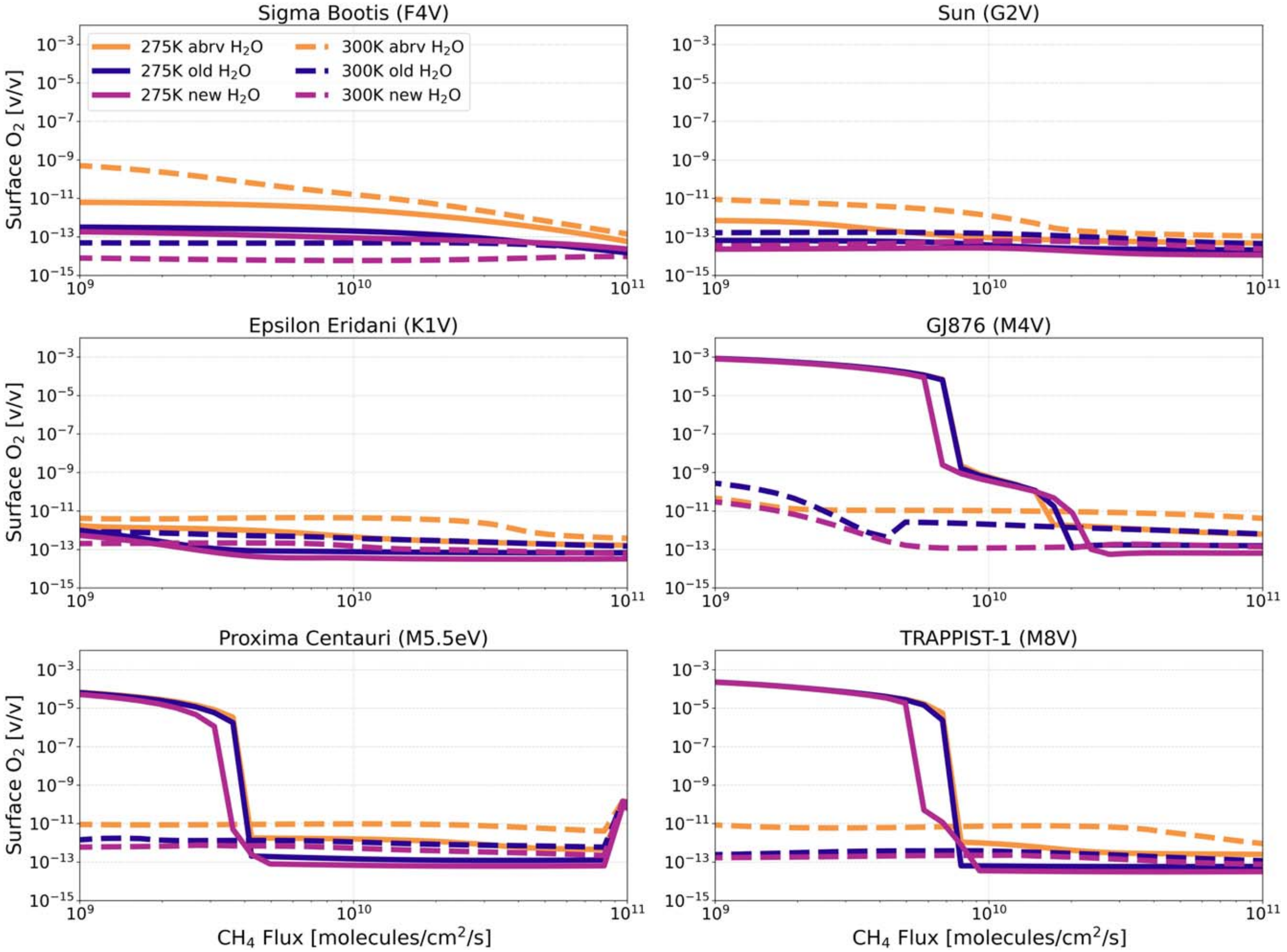

**Figure 5.** $H_2O$ cross-section sensitivity test; surface $O_2$ vs. $CH_4$ flux for anoxic habitable planets orbiting FGKM-type host stars.

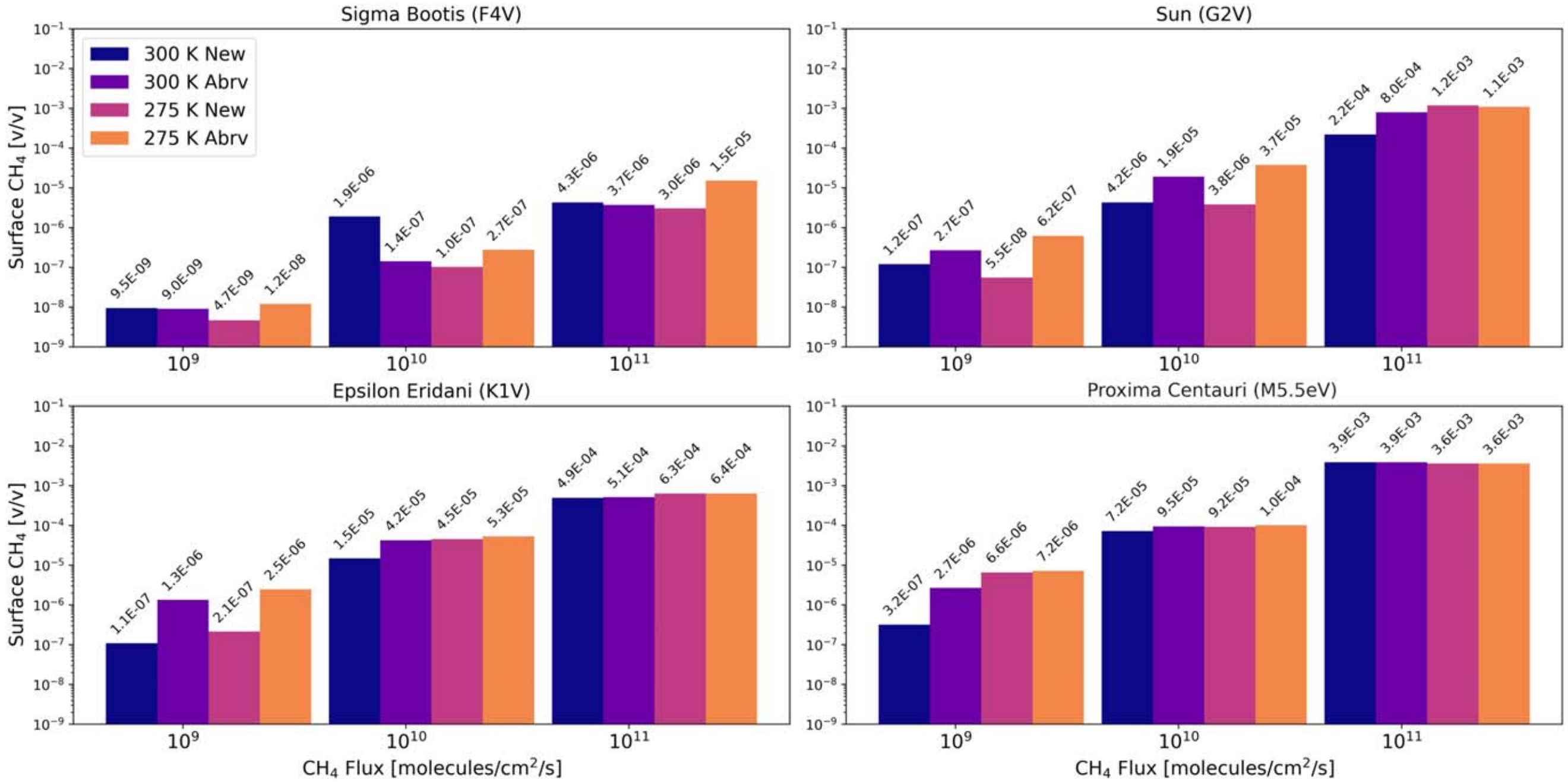

**Figure 6.** Quantitative difference between $CH_4$ surface mixing ratios for $CH_4$ surfaces fluxes of $10^9$, $10^{10}$, and $10^{11}$ molecules $cm^{-2}$ $s^{-1}$, for anoxic habitable planets modeled using Sigma Bootis, the Sun, Epsilon Eridani, and Proxima Centauri as the host star.

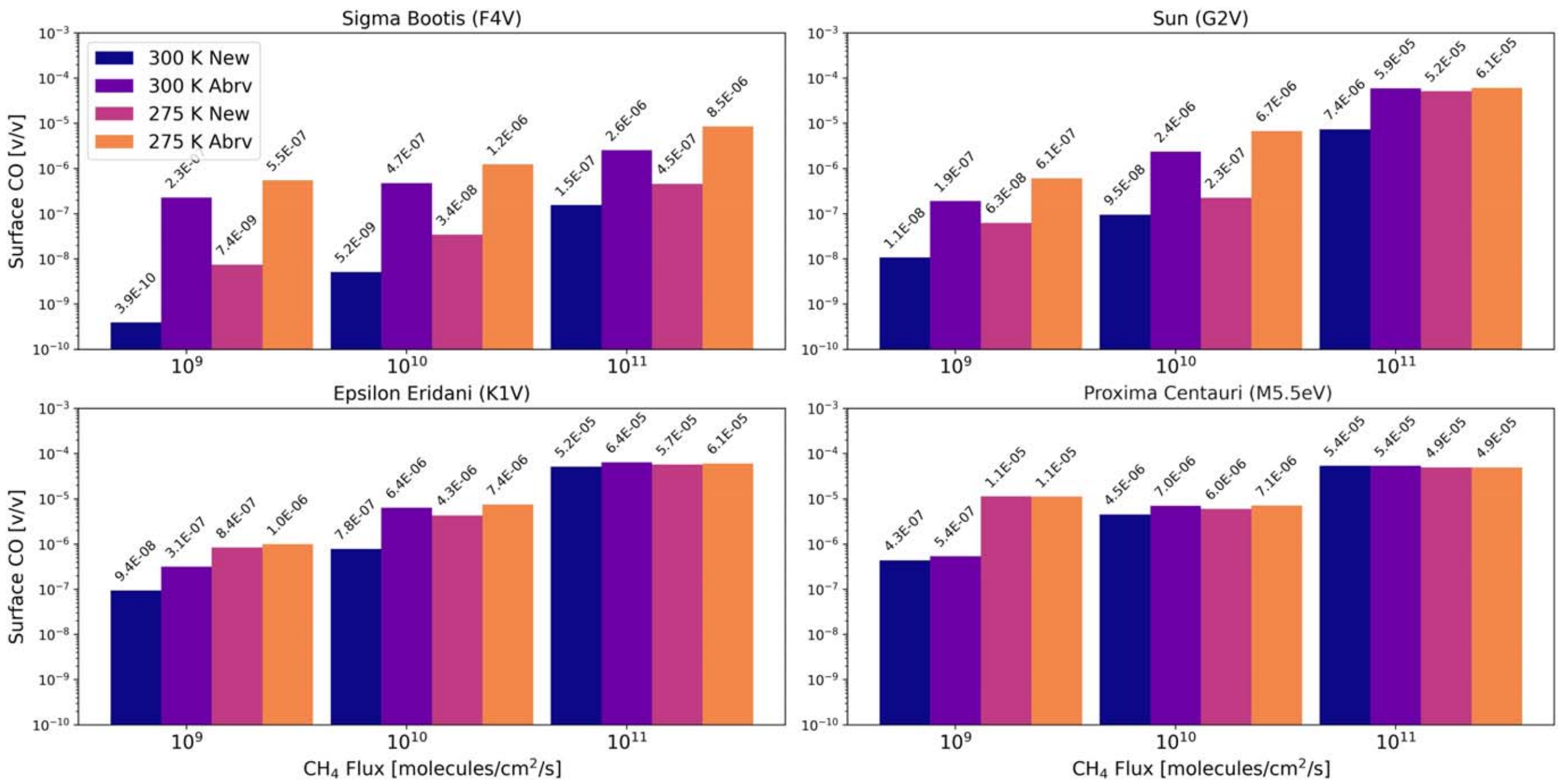

**Figure 7.** Quantitative difference between CO surface mixing ratios for $CH_4$ surfaces fluxes of $10^9$, $10^{10}$, and $10^{11}$ molecules $cm^{-2}$ $s^{-1}$, for anoxic habitable planets modeled using Sigma Bootis, the Sun, Epsilon Eridani, and Proxima Centauri as the host star.

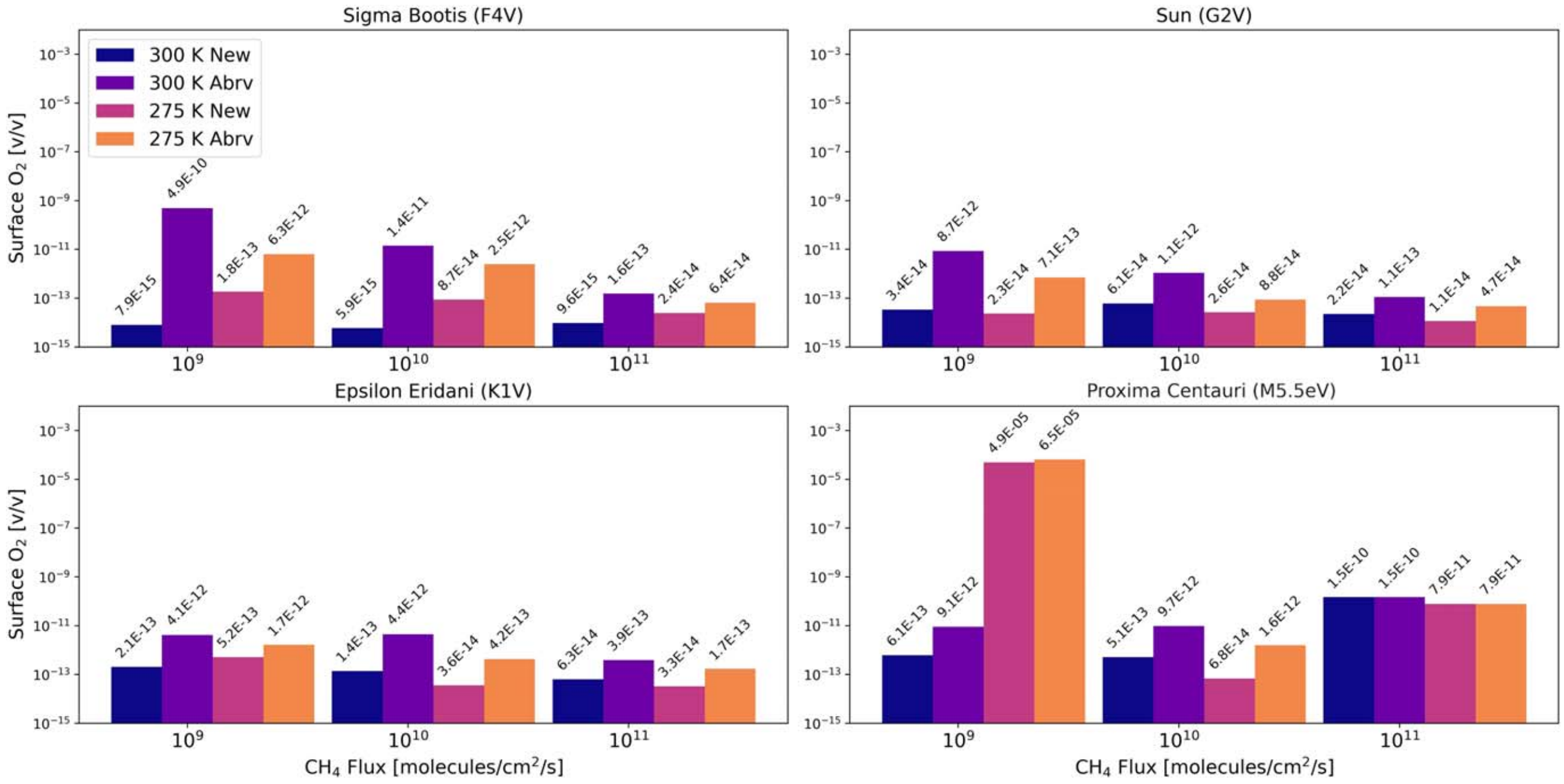

**Figure 8.** Quantitative difference between $O_2$ surface mixing ratios for $CH_4$ surfaces fluxes of $10^9$, $10^{10}$, and $10^{11}$ molecules $cm^{-2}$ $s^{-1}$, for anoxic habitable planets modeled using Sigma Bootis, the Sun, Epsilon Eridani, and Proxima Centauri as the host star.

Although we see the largest order of magnitude variations between $CH_4$ levels for planets orbiting the Sun with a $CH_4$ flux of $1.1 \times 10^{10}$ molecules $cm^{-2}$ $s^{-1}$, the largest absolute differences occur at higher fluxes. These larger absolute differences would not necessarily translate to bigger differences in the observed spectral features because absorption bands can become saturated at higher abundances. Examining the spectral variation of $CH_4$ features at each $CH_4$ flux due to differing $H_2O$ cross sections is beyond the scope of this paper, which is meant to demonstrate that such inputs can have a substantial impact in some regions of flux-abundance space and that they are therefore worth characterizing accurately and precisely.

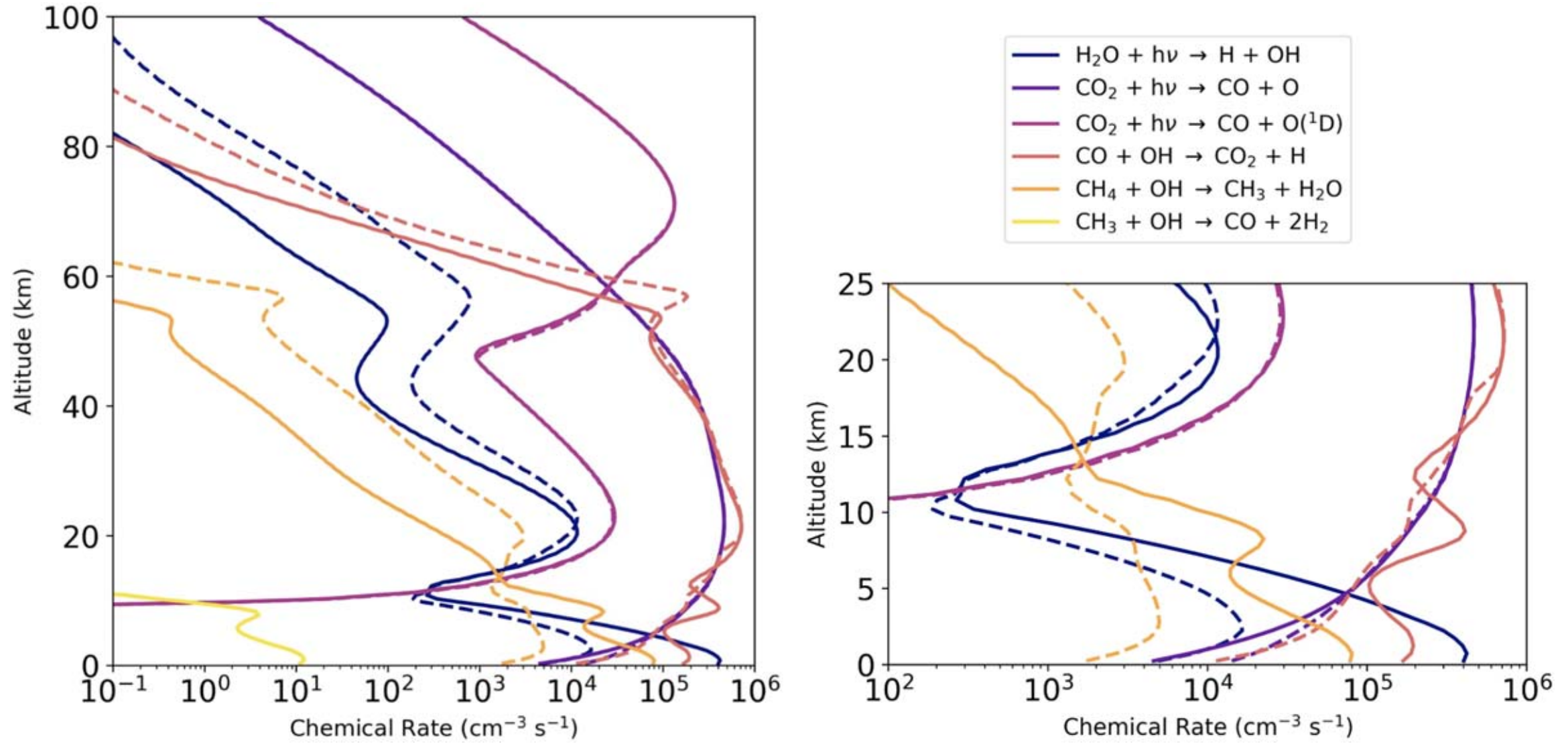

**Figure 9.** Reaction rates for an anoxic habitable planet orbiting the Sun, $CH_4 = 1.1 \times 10^{10}$ molecules $cm^{-2}$ $s^{-1}$, for the 275 K temperature regime. Solid lines represent reaction rates modeled using the new $H_2O$ cross sections, dashed lines represent reaction rates modeled using the abbreviated $H_2O$ cross sections. The plot on the left shows the reaction rates up to 100 km; on the right, the rates up to 25 km are shown.

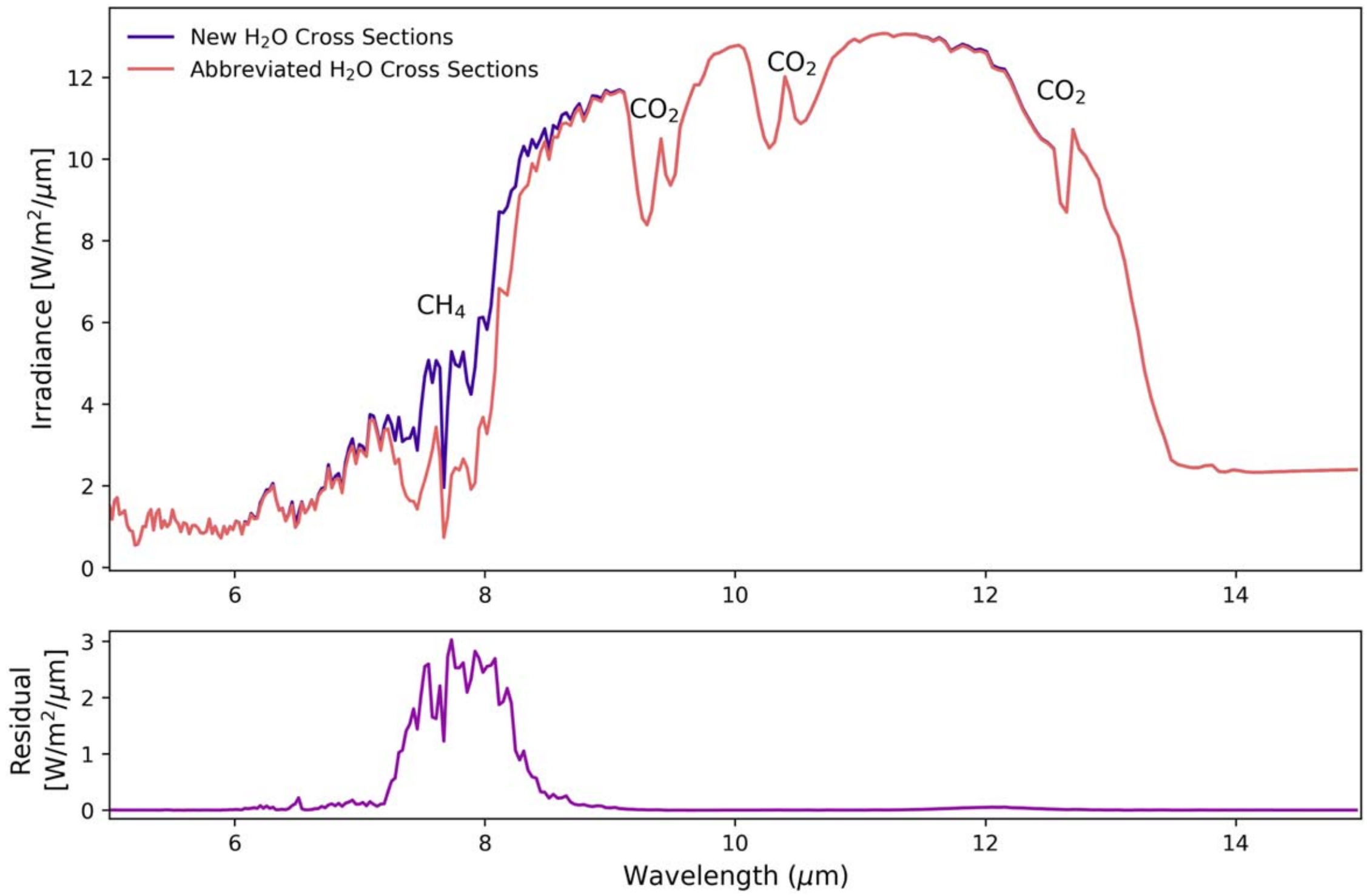

**Figure 10.** Top: emission spectra comparison between the new and abbreviated cross sections, using the Sun as the host star, where the $CH_4$ flux $= 1.10 \times 10^{10}$ molecules $cm^{-2}$ $s^{-1}$. Bottom: plot of the residual, obtained by subtracting the abbreviated $H_2O$ cross-section spectrum from the new $H_2O$ cross-section spectrum.

## 4. Discussion

The results presented in Section 3 are promising for JWST M-dwarf observation targets, with the M-type host stars showing the least variation with respect to $H_2O$ cross-section prescription. As planets orbiting M-stars are the primary targets of JWST for atmospheric characterization, this means interpreting results from JWST should be less affected by the choice of $H_2O$ cross section in the UV. $H_2O$ cross-section prescription had the greatest impact on the higher-mass host stars; these stars are brighter in the MUV than the lower-mass M-type stars (see Figure 1). FGK-type stars will be the targets of the upcoming HWO, which will be optimized for observing reflected light from small planets in the IR, UV, and optical (Mamajek & Stapelfeldt 2024). Our results show that for these higher-mass host stars, having accurate $H_2O$ cross sections will be critical for making accurate predictions and correct interpretations of temperate terrestrial exoplanet spectra.

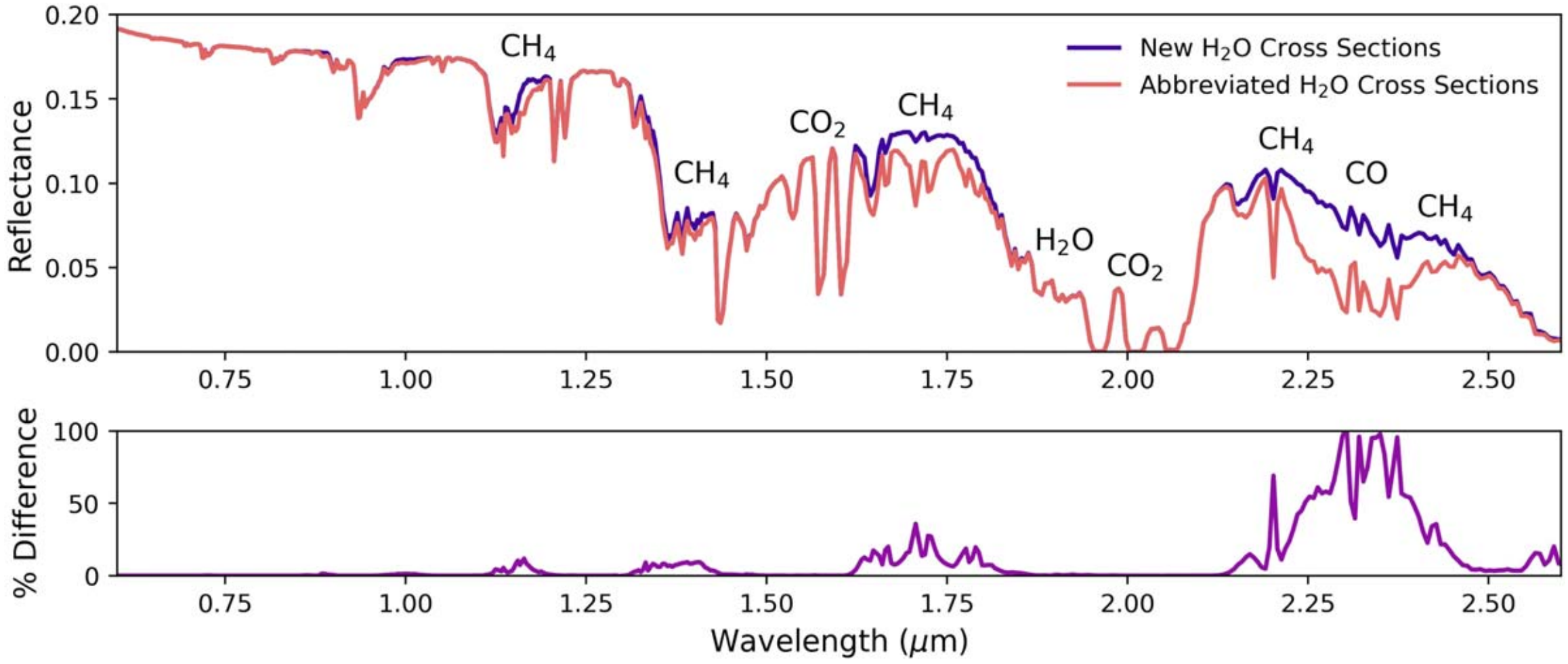

**Figure 11.** Top: reflection spectra comparison between the new and abbreviated cross sections, using the Sun as the host star, where the $CH_4$ flux = $1.10 \times 10^{10}$ molecules $cm^{-2}$ $s^{-1}$. Bottom: plot of the percent difference between the abbreviated $H_2O$ cross-section spectrum and the new $H_2O$ cross-section spectrum.

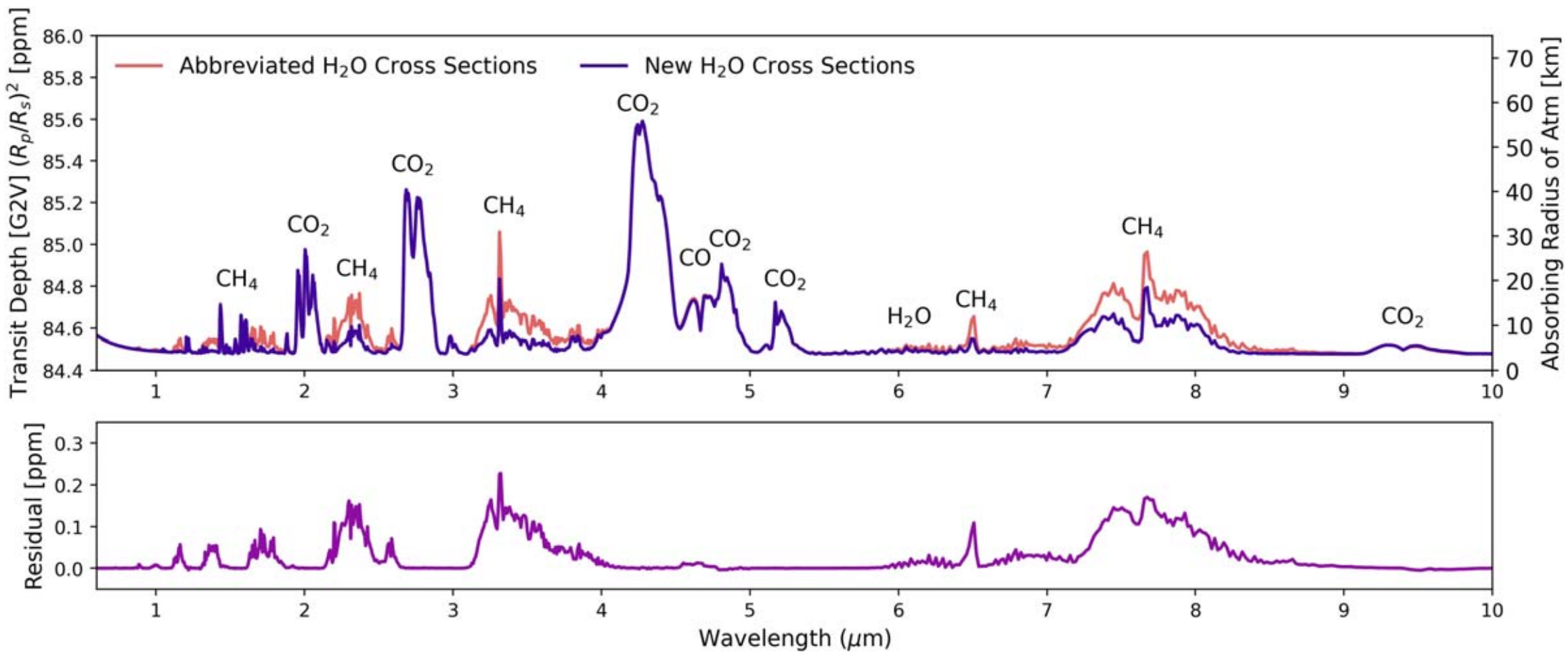

**Figure 12.** Top: transmission spectra comparison between the new and abbreviated cross sections, using the Sun as the host star, where the $CH_4$ flux = $1.10 \times 10^{10}$ molecules $cm^{-2}$ $s^{-1}$. Bottom: plot of the residual, obtained by subtracting the abbreviated $H_2O$ cross-section spectrum from the new $H_2O$ cross-section spectrum.

We have tested only an $N_2$–$CO_2$–$H_2O$ dominated atmosphere, with a composition similar to that of the Archean Earth. Different atmospheric compositions will be more or less affected by the updated $H_2O$ cross sections; for instance, an oxygen-rich atmosphere with a stratospheric ozone layer that shields water in the troposphere from MUV photons and/or where OH is not primarily sourced from $H_2O$ photolysis (e.g., the modern Earth atmosphere) will not be as sensitive to $H_2O$ cross sections. On the other hand, in our simulations, we keep the surface $CO_2$ volume mixing ratio fixed at 2%. As demonstrated for Sun-like host stars in Watanabe & Ozaki (2024), higher relative abundances of $CO_2$ naturally lead to higher rates of $CO_2$ photolysis and further suppression of $H_2O$ photolysis. The MUV $H_2O$ cross sections would become increasingly important in high-$CO_2$ cases, owing to the increased CO sourced from $CO_2$ photolysis combined with a reduction in OH produced from FUV $H_2O$ photolysis (Akahori et al. 2023).

The updated cross sections are also a conservative estimate, meaning the predicted $H_2O$ photolysis occurring with these cross sections is a lower estimate and actual photolysis rates may be higher (Ranjan et al. 2020). Thus, OH production may be greater in magnitude, and subsequent atmospheric trace gas abundances may be even lower in abundance than what is predicted here (e.g., Appendix C, where we see that using the cutoff prescription from Ranjan et al. 2020 results in around half the amount of surface CO and $CH_4$ as is generated with the extrapolation prescription). These updated $H_2O$ cross sections differ in the FUV somewhat from previous recommendations (see Figure 1, specifically around 180 nm and 120–140 nm); to have the best cross-section recommendations possible, it is important to further investigate these differences in the FUV range. Further, the MUV cross sections are the result of absorption from (thermally populated) $H_2O$ rotational levels in the first $H_2O$ vibrational level. The cross sections we use for this long-wavelength tail have been measured at one temperature (292 K), but the temperature-dependent behavior of the cross sections could matter. The effect is likely small at the temperature regimes we have simulated, but discrepancies may result if we wanted to use these MUV cross sections for temperature regimes substantially different from habitable temperatures, especially high temperatures (e.g., steam atmospheres). Ideally, precise temperature-dependent cross sections should be measured (or calculated) and adopted. We have not tested the sensitivity of these updated cross sections to chemical reaction rates; reaction rates previously identified as needing further study (e.g., the reactions identified in Ranjan et al. 2020) may yet lead to further impacts on atmospheric trace gas abundances. Additionally, the $CO_2$ cross sections used in these simulations also suffer from uncertainty, and it is important to test the sensitivity of the $H_2O$ cross-section prescription to the choice of $CO_2$ cross-section prescription in order to examine competing effects.

Similar to the conclusions in Watanabe & Ozaki (2024), our results show that CO runaway occurs more easily for planets orbiting the G- and K-type stars at the colder temperature regimes, and does not occur for the F-type star even at the highest $CH_4$ fluxes. For the M-type host stars at low $CH_4$ fluxes in the 275 K surface temperature regime, we see moderate CO and $O_2$ surface mixing ratios (Figure 5). It is important to note that this was only seen in the biotic scenario, where CO is being deposited on the surface at a rate of $1.2 \times 10^{-4}$ cm s$^{-1}$, and not in the abiotic scenario (see Appendix B, Figure 15).

We have modeled the impact on three atmospheric gases: $CH_4$, CO, and trace $O_2$. These are not the only gases that would be affected by the choice of $H_2O$ cross sections. Any gas that interacts with OH (e.g., phosphine, ethane, methyl chloride, to name a few) will be affected by the increased OH produced by extended $H_2O$ cross sections. Thus, further constraining and testing relationships of other trace gases to these $H_2O$ inputs will be vital for diverse planetary biosignature scenarios (e.g., Domagal-Goldman et al. 2011; Sousa-Silva et al. 2020; Madhusudhan et al. 2021; Huang et al. 2022; Leung et al. 2022). Our results show that the updated $H_2O$ cross sections are essential for the interpretation of future observational data. While the updated cross sections have unprecedented accuracy compared to previous results, additional sensitive experimental results could help distinguish between the conservative extrapolation prescription and the possible higher MUV opacity shown by the measurements. Our results presented in Appendix C show meaningful differences in predicted gas mixing ratios between the extrapolation and *cutoff* $H_2O$ prescriptions presented in Ranjan et al. (2020), which are both conservative prescriptions in terms of their treatment of photodissociation opacity at wavelengths >216 nm. Therefore future, more sensitive measurements of $H_2O$ MUV opacities are justified.

It is also important to consider downstream effects we have not modeled, which may or may not be spectrally observable but will nonetheless have consequences for atmospheric chemistry. For instance, organic haze can start to form at high $CH_4/CO_2$ ratios (Trainer et al. 2006; Arney et al. 2016, 2018). Given that the updated $H_2O$ cross sections lead to lower $CH_4$ mixing ratios, a greater $CH_4$ flux may be required in order to form haze. Further investigation into the effects of these updated cross sections on haze formation is an important topic for future work.

## 5. Conclusions

Having accurate fundamental inputs, such as photochemical cross sections, is critical for modeling and interpreting the atmospheres of temperate terrestrial exoplanets. Depending on which $H_2O$ cross-section prescription is used, and what wavelength is used as the terminating wavelength, we predict meaningful differences in resulting trace gas abundances. Temperate anoxic planets orbiting higher-mass FGK-type host stars see the largest variations, with trace gas abundances varying by several orders of magnitude depending on the temperature regime, host star, and cross-section prescription. Moreover, these variations can become significant at threshold $CH_4$ fluxes, which straddle the line between a reasonable abiotic flux and a potentially biotic flux. Depending on the capabilities and noise thresholds of upcoming observatories, these orders of magnitude variations in the $CH_4$ volume mixing ratios may cause an observable difference in the resulting planetary reflection, emission, and transmission spectra.

The relatively small variations we observe for M-type host stars are a positive result for interpretations of JWST observations of secondary atmospheres—e.g., the $H_2O$ cross-section prescription should not introduce significant uncertainty in interpreting these observations. In contrast, our results indicate that accurate $H_2O$ cross sections will be critical for robust interpretation of observations from HWO, as the chosen prescription can have a significant impact on the modeled atmospheres and simulated spectra for planets orbiting FGK-type stars. Finally, these results also reemphasize the general need to continue investigating fundamental uncertainty in the basic inputs underlying photochemical models, and highlight the critical importance of obtaining empirical cross sections to accurately interpret exoplanet observations.

## Acknowledgments

This work was performed by the Experimental Constraints for Improving Terrestrial Exoplanet Photochemical Models (ExCITE-PM) Team funded by NASA Exoplanet Research Program (XRP) grant No. 80NSSC22K0235. We thank the anonymous reviewer for the feedback that helped to improve this paper. E.W.S., W.B., and C.T.R. acknowledge further support from the NASA Interdisciplinary Consortium for Astrobiology Research (ICAR) with funding issued through the Alternative Earths Team (grant No. 80NSSC21K0594) and the Consortium on Habitability and Atmospheres of M-dwarf Planets (CHAMPS) Team (grant No. 80NSSC21K0905). CSS acknowledges additional support from the Heising-Simons Foundation.

*Software:* Atmos (Arney et al. 2016), SMART (Meadows & Crisp 1996; Crisp 1997)

## Appendix A
## Biotic Boundary Conditions

Table 2 presents the boundary conditions used for the biotic scenario. Each species boundary condition is specified as a surface flux, a surface mixing ratio, a dry deposition velocity, or as a combination of a surface flux and dry deposition velocity.

**Table 2**
Atmospheric Species Boundary Conditions

| Species | Surface Flux (molecules $cm^{-2}$ $s^{-1}$) | Surface Mixing Ratio (v/v) | Dry Deposition Velocity (cm $s^{-1}$) |
|---|---|---|---|
| O | ⋯ | ⋯ | 1 |
| $O_2$ | ⋯ | ⋯ | 0 |
| $H_2O$ | ⋯ | ⋯ | 0 |
| H | ⋯ | ⋯ | 1 |
| OH | ⋯ | ⋯ | 1 |
| $HO_2$ | ⋯ | ⋯ | 1 |
| $H_2O_2$ | ⋯ | ⋯ | $2 \times 10^{-1}$ |
| $H_2$ | $1 \times 10^{10}$ | ⋯ | $2.4 \times 10^{-4}$ |
| CO | $1 \times 10^{8}$ | ⋯ | $1.2 \times 10^{-4}$ |
| HCO | ⋯ | ⋯ | 1 |
| $H_2CO$ | ⋯ | ⋯ | $2 \times 10^{-1}$ |
| $CH_4$ | $10^{9}$ - $10^{11}$ | ⋯ | ⋯ |
| $CH_3$ | ⋯ | ⋯ | 1 |
| $C_2H_6$ | ⋯ | ⋯ | 0 |
| NO | ⋯ | ⋯ | $3 \times 10^{-4}$ |
| $NO_2$ | ⋯ | ⋯ | $3 \times 10^{-3}$ |
| HNO | ⋯ | ⋯ | 1 |
| $O_3$ | ⋯ | ⋯ | $7 \times 10^{-2}$ |
| $HNO_3$ | ⋯ | ⋯ | $2 \times 10^{-1}$ |
| N | ⋯ | ⋯ | 0 |
| $C_3H_2$ | ⋯ | ⋯ | 0 |
| $C_3H_3$ | ⋯ | ⋯ | 0 |
| $CH_3C_2H$ | ⋯ | ⋯ | 0 |
| $CH_2CCH_2$ | ⋯ | ⋯ | 0 |
| $C_3H_5$ | ⋯ | ⋯ | 0 |
| $C_3H_6$ | ⋯ | ⋯ | 0 |
| $C_3H_7$ | ⋯ | ⋯ | 0 |
| $C_3H_8$ | ⋯ | ⋯ | 0 |
| $C_2H_4OH$ | ⋯ | ⋯ | 0 |
| $C_2H_2OH$ | ⋯ | ⋯ | 0 |
| $C_2H_5$ | ⋯ | ⋯ | 0 |
| $C_2H_4$ | ⋯ | ⋯ | 0 |
| CH | ⋯ | ⋯ | 0 |
| $CH_3O_2$ | ⋯ | ⋯ | 0 |
| $CH_3O$ | ⋯ | ⋯ | 0 |
| $CH_2CO$ | ⋯ | ⋯ | 0 |
| $CH_3CO$ | ⋯ | ⋯ | 0 |
| $CH_3CHO$ | ⋯ | ⋯ | 0 |
| $C_2H_2$ | ⋯ | ⋯ | 0 |
| $CH_23$ | ⋯ | ⋯ | 0 |
| $C_2H$ | ⋯ | ⋯ | 0 |
| $C_2$ | ⋯ | ⋯ | 0 |
| $C_2H_3$ | ⋯ | ⋯ | 0 |
| HCS | ⋯ | ⋯ | 0 |
| $CS_2$ | ⋯ | ⋯ | 0 |
| CS | ⋯ | ⋯ | 0 |
| OCS | ⋯ | ⋯ | 0 |
| S | ⋯ | ⋯ | 0 |
| HS | ⋯ | ⋯ | 0 |
| $H_2S$ | $3.5 \times 10^{8}$ | ⋯ | $2 \times 10^{-2}$ |
| $SO_3$ | ⋯ | ⋯ | 0 |
| HSO | ⋯ | ⋯ | 1 |
| $H_2SO_4$ | ⋯ | ⋯ | 1 |
| $SO_2$ | $3 \times 10^{9}$ | ⋯ | 1 |
| SO | ⋯ | ⋯ | 0 |
| $CO_2$ | ⋯ | $2 \times 10^{-2}$ | ⋯ |
| $SO_4$ AER | ⋯ | ⋯ | 0.01 |
| $S_8$ AER | ⋯ | ⋯ | 0.01 |
| HCAER | ⋯ | ⋯ | 0.01 |
| HCAER2 | ⋯ | ⋯ | 0.01 |

## Appendix B
## Abiotic Scenario

The $CH_4$ flux sensitivity tests were also conducted for an abiotic scenario; the results of these additional tests are shown in Figures 13–15. The key change from the biotic scenario is the deposition velocity of CO. In the biotic scenario, the deposition velocity of CO is set to $1.2 \times 10^{-4}$ cm $s^{-1}$; in the abiotic scenario, this deposition velocity is only $1.0 \times 10^{-8}$ cm $s^{-1}$. This decreased CO deposition velocity allows CO to build up more easily in the atmosphere, saturating the OH sink. Because of this, the abiotic results are generally less sensitive to the $H_2O$ cross section. It is plausible that a given planetary scenario could include habitable planets with CO deposition rates between the biotic and abiotic limits we have used in these two scenarios. Thus, these two scenarios serve more generally as upper and lower limits of how the $H_2O$ cross-section prescriptions could impact planetary atmospheres.

For this abiotic scenario, we also conducted a limited sensitivity test to investigate the impact of enforcing a global redox balance. The redox balance hypothesis is often assumed, but is as of yet unproven. To conduct this test, we used the G-type host star with the 275 K surface temperature and a $CH_4$ surface flux of $1.0 \times 10^{10}$ molecules $cm^{-2}$ $s^{-1}$. We adjusted the $H_2$ surface flux until global redox was attained, which occurred at a $H_2$ flux of around $4.4 \times 10^{11}$ molecules $cm^{-2}$ $s^{-1}$. With global redox imposed, we found the difference between the surface $CH_4$ volume mixing ratios for the new and abbreviated $H_2O$ cross sections was minimized (though they still differed by about 5%). The resulting surface CO mixing ratios were still larger for the abbreviated cross sections by about an order of magnitude. These results are consistent with the results from Ranjan et al. (2020), who showed that assuming global redox balance decreased the impact of the $H_2O$ cross-section prescription to surface $CH_4$, but the impact on surface CO remained.

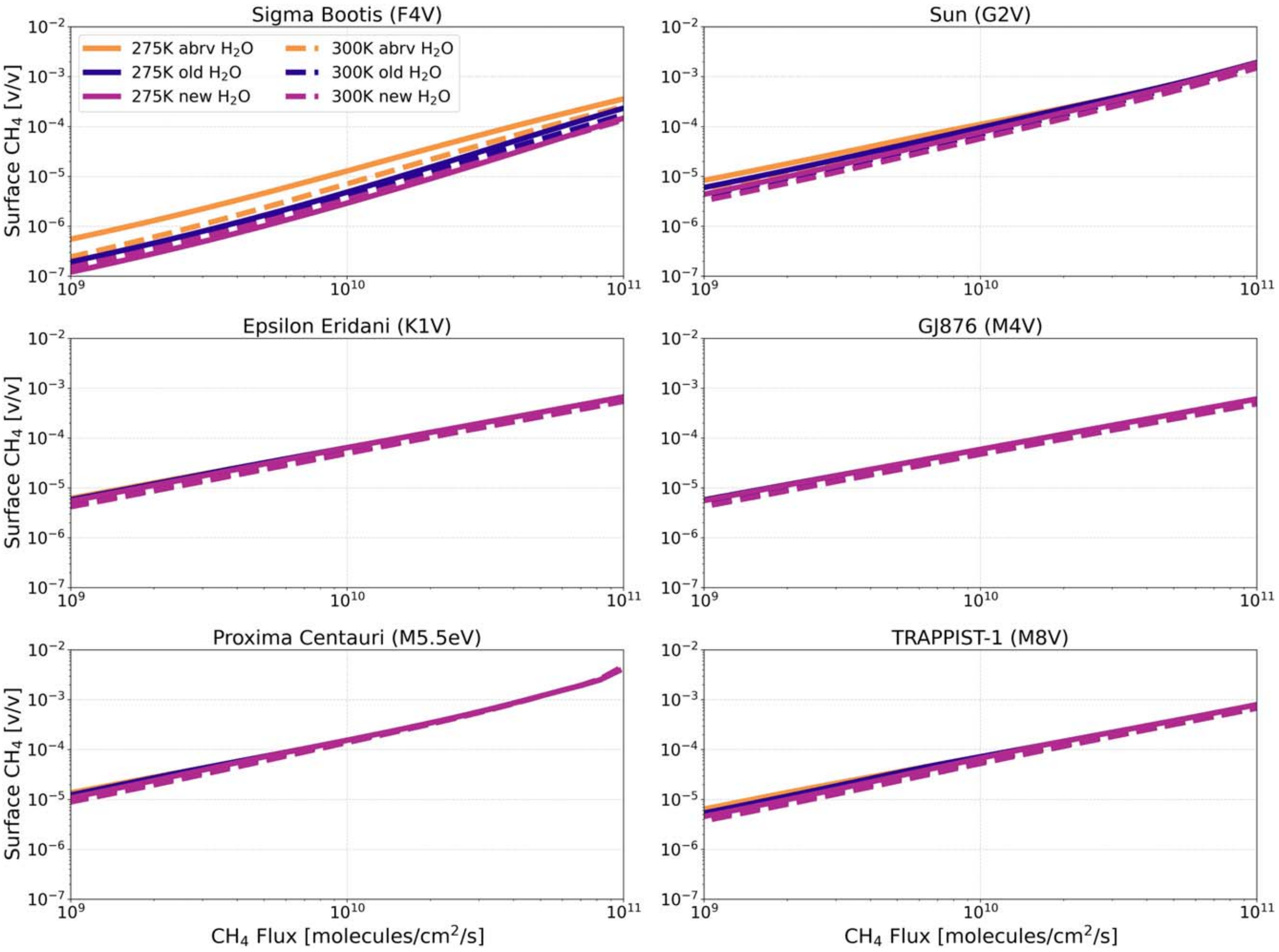

**Figure 13.** Abiotic $H_2O$ cross-section sensitivity test; surface $CH_4$ vs. $CH_4$ flux for planets orbiting FGKM-type host stars.

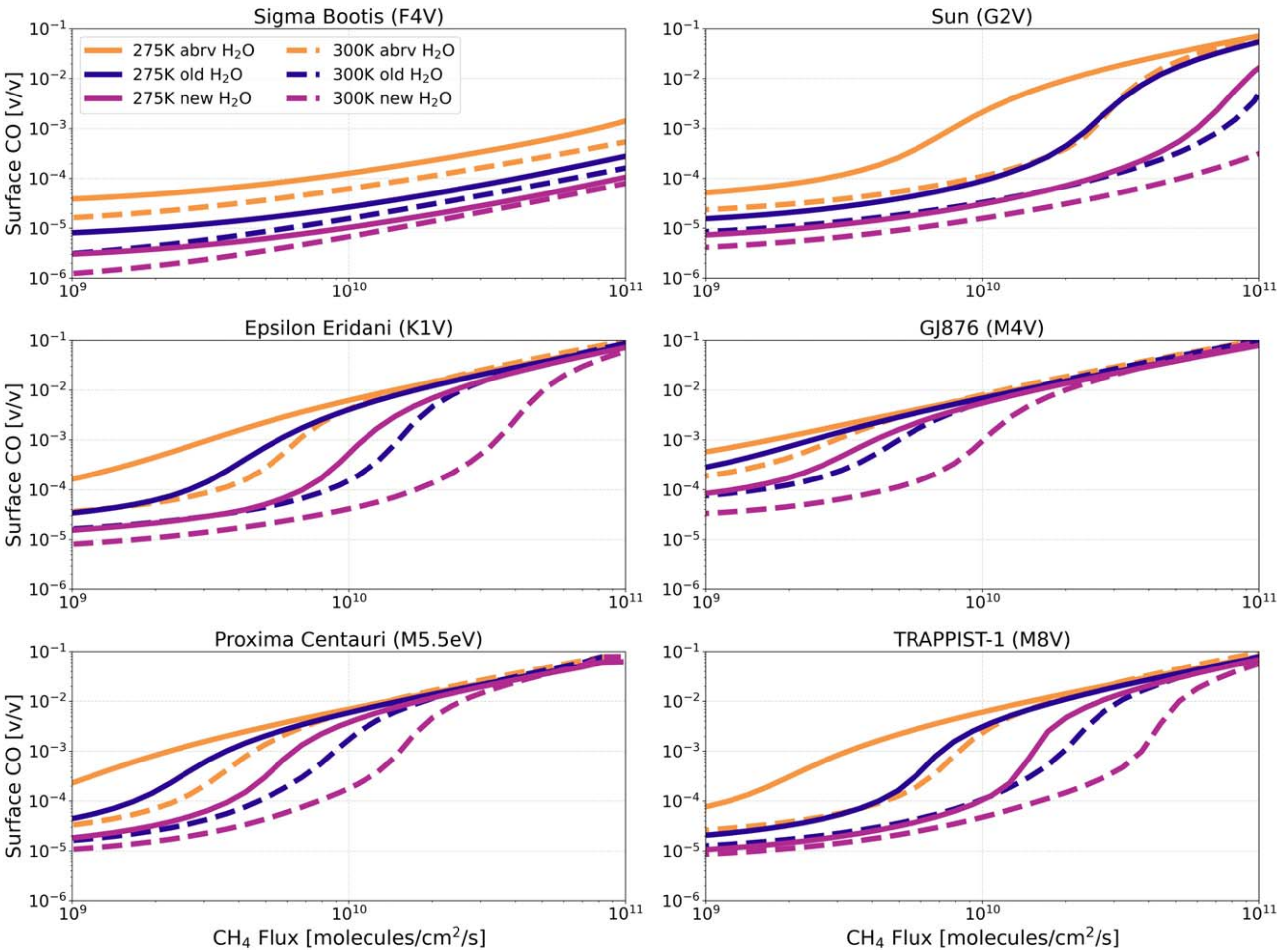

**Figure 14.** Abiotic $H_2O$ cross-section sensitivity test; surface CO vs. $CH_4$ flux for planets orbiting FGKM-type host stars.

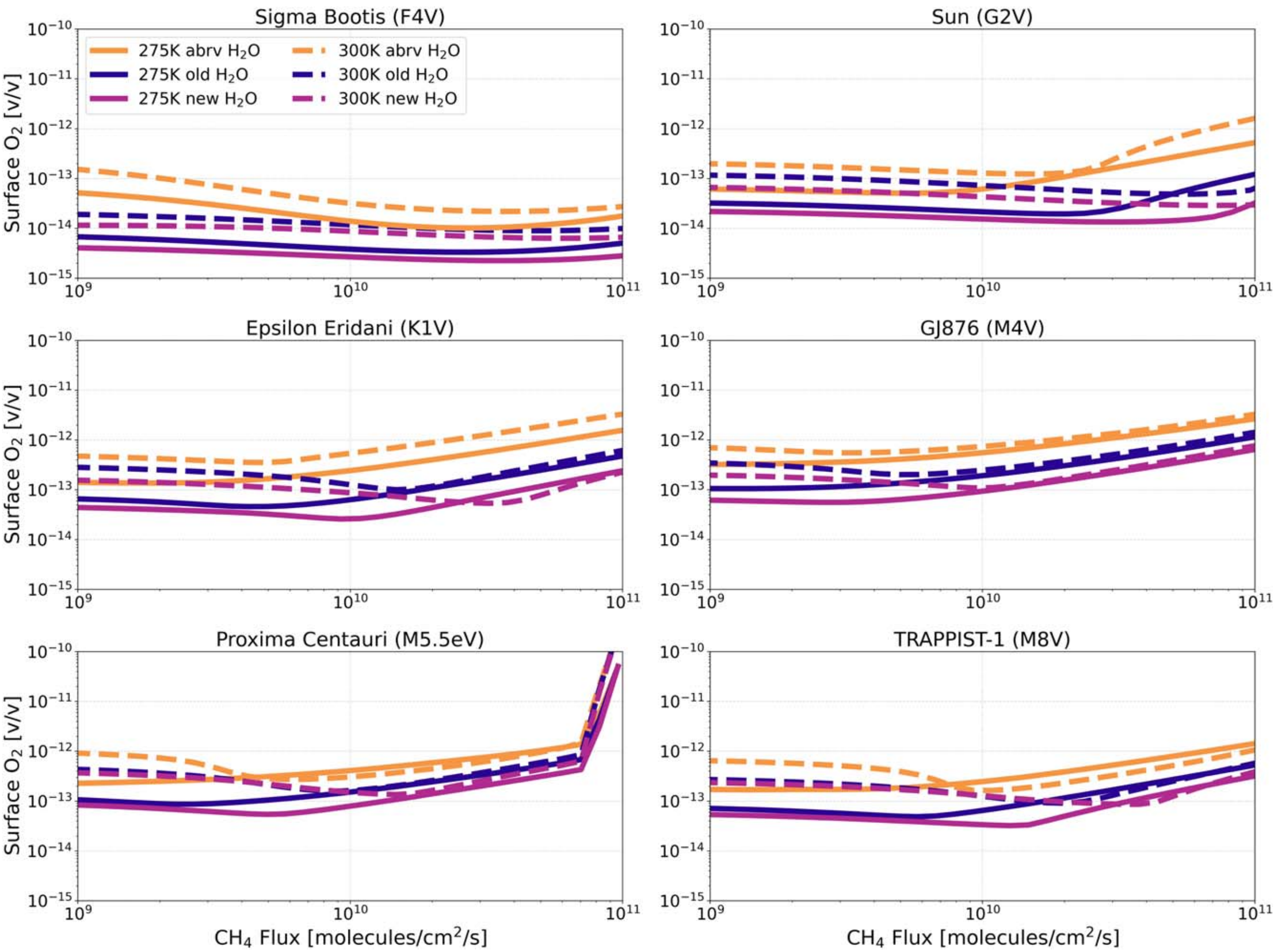

**Figure 15.** Abiotic $H_2O$ cross-section sensitivity test; surface $O_2$ vs. $CH_4$ flux for planets orbiting FGKM-type host stars.

## Appendix C
## Cutoff versus Extrapolated Cross-section Prescriptions

Figure 16 is a modified version of Figure 1, which shows the $H_2O$ cross sections as a function of wavelength, now including the cutoff prescription from Ranjan et al. (2020), which terminates around 216 nm. Figure 17 shows the results for $H_2O$ cross-section sensitivity tests using the Sun as a host star, for surface CO and $CH_4$ as a function of $CH_4$ surface flux, including the cutoff cross-section prescription.

Although the resulting surface CO and $CH_4$ volume mixing ratios do not differ by orders of magnitude between the cutoff prescription and the extrapolation prescription, there are still noticeable differences. For CO, the greatest fractional differences occur around $3.22 \times 10^{10}$ and $2.36 \times 10^{10}$ molecules $cm^{-2}$ $s^{-1}$ for the 275 and 300 K temperature regimes, respectively, where the extrapolated cross sections result in 2.3 and 2.0 times the amount of surface CO as with the cutoff cross sections. For $CH_4$, the greatest fractional differences occur around $3.22 \times 10^{10}$ and $9.61 \times 10^{10}$ molecules $cm^{-2}$ $s^{-1}$ for the 275 and 300 K temperature regimes, respectively, where the extrapolated cross sections result in 1.8 and 1.5 times the amount of surface $CH_4$ as with the cutoff cross sections.

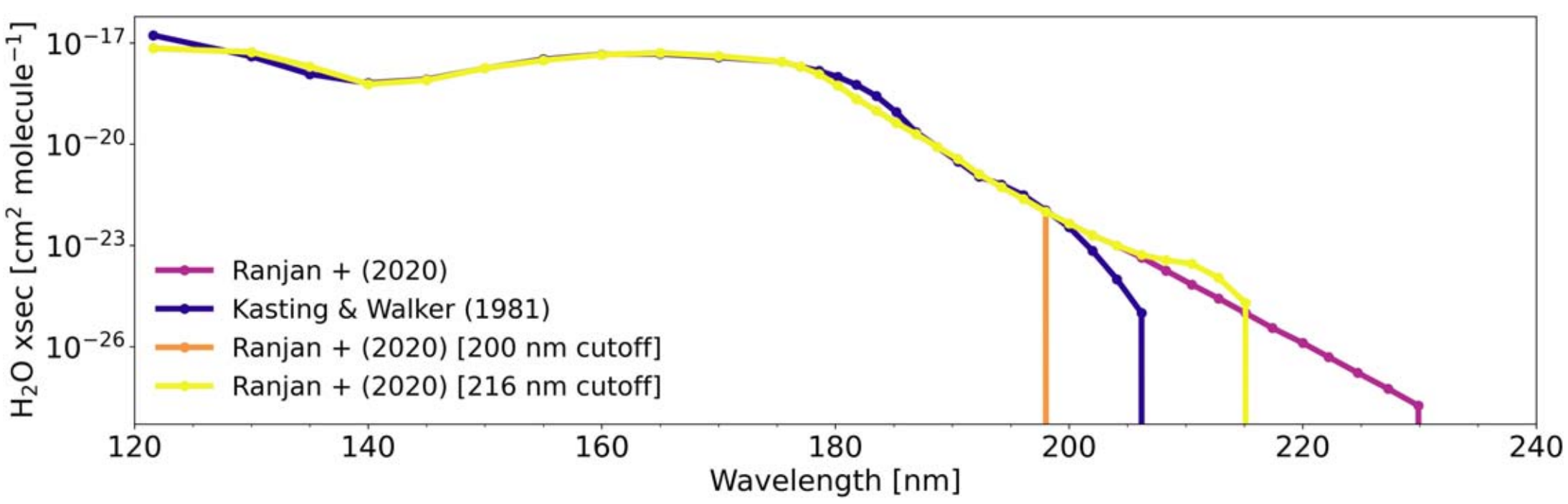

**Figure 16.** $H_2O$ cross-section prescriptions from 120–240 nm, including the cutoff prescription.

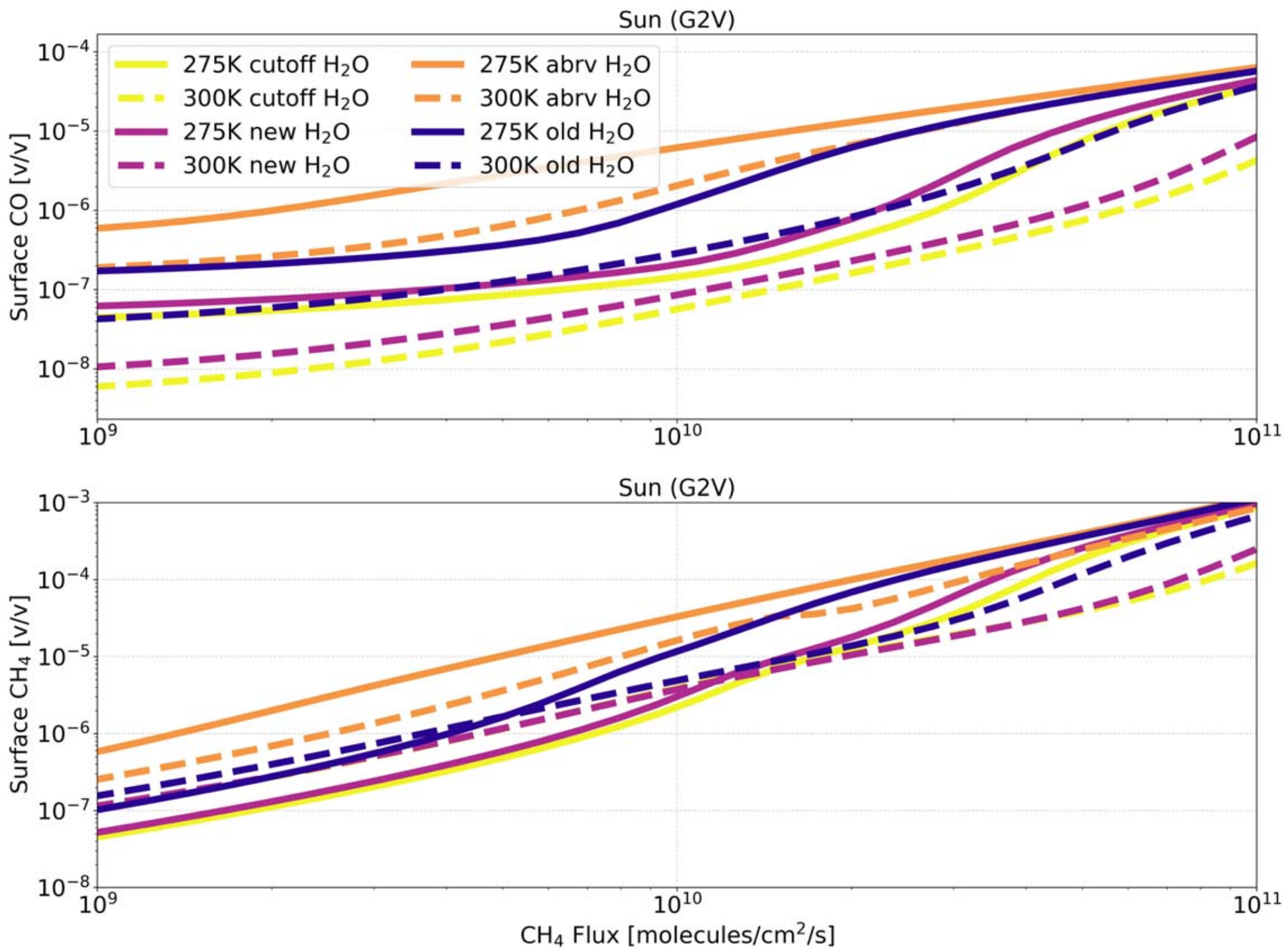

**Figure 17.** Biotic $H_2O$ cross-section sensitivity test, including the cutoff prescription, using the Sun as a host star. (Top): surface CO vs. $CH_4$ flux. (Bottom): surface $CH_4$ vs. $CH_4$ flux.

## Appendix D
## Column Density Relationships

Because the reflected and emitted light spectra of planetary atmospheres are dependent on the column density of a gas, rather than its surface mixing ratio, we reproduce Figures 3–5 by showing the column density as the dependent variable. Importantly, we show that the differences in the $CH_4$ column densities from $H_2O$ cross-section prescription (Figure 18) closely follow the same relationships shown in Figure 3 for $CH_4$ surface mixing ratios. This is because, as seen in Figure 2, the $CH_4$ vertical profiles run parallel, keeping the same difference regardless of the altitude.

On the other hand, CO shows the greatest differences at the surface; above about 20 km, the difference between the vertical profiles goes to zero. Since most of the atmospheric CO is present above 20 km, Figure 19, showing CO column density versus $CH_4$ flux, exhibits a different relationship from Figure 4. This is most apparent with the F-type host star, which has a CO column density from 1–4 $\times$ $10^{20}$ molecules cm$^{-2}$ regardless of the surface $CH_4$ flux or $H_2O$ cross-section prescription. For this host star, the increased $CH_4$ flux leads to more $H_2O$ production through the reaction of the reaction OH + $CH_4$ $\rightarrow$$CH_3$ + $H_2O$. The $CH_3$ produced in this reaction can further interact with OH in the reaction OH + $CH_3$ $\rightarrow$ CO + $2H_2$. Likewise, the additional $H_2O$ in the upper atmosphere can undergo photolysis, producing more OH, which then consumes the CO. Because these reactions are occurring higher up in the atmosphere, differences between $H_2O$ cross-section prescriptions in the MUV have less of an impact.

For the $O_2$ column density shown in Figure 20, we do see a similar relationship as in Figure 5, though to a smaller degree. This is because, as with CO, most of the $O_2$ is present at more significant levels above about 20 km, and smaller changes at the surface become less apparent.

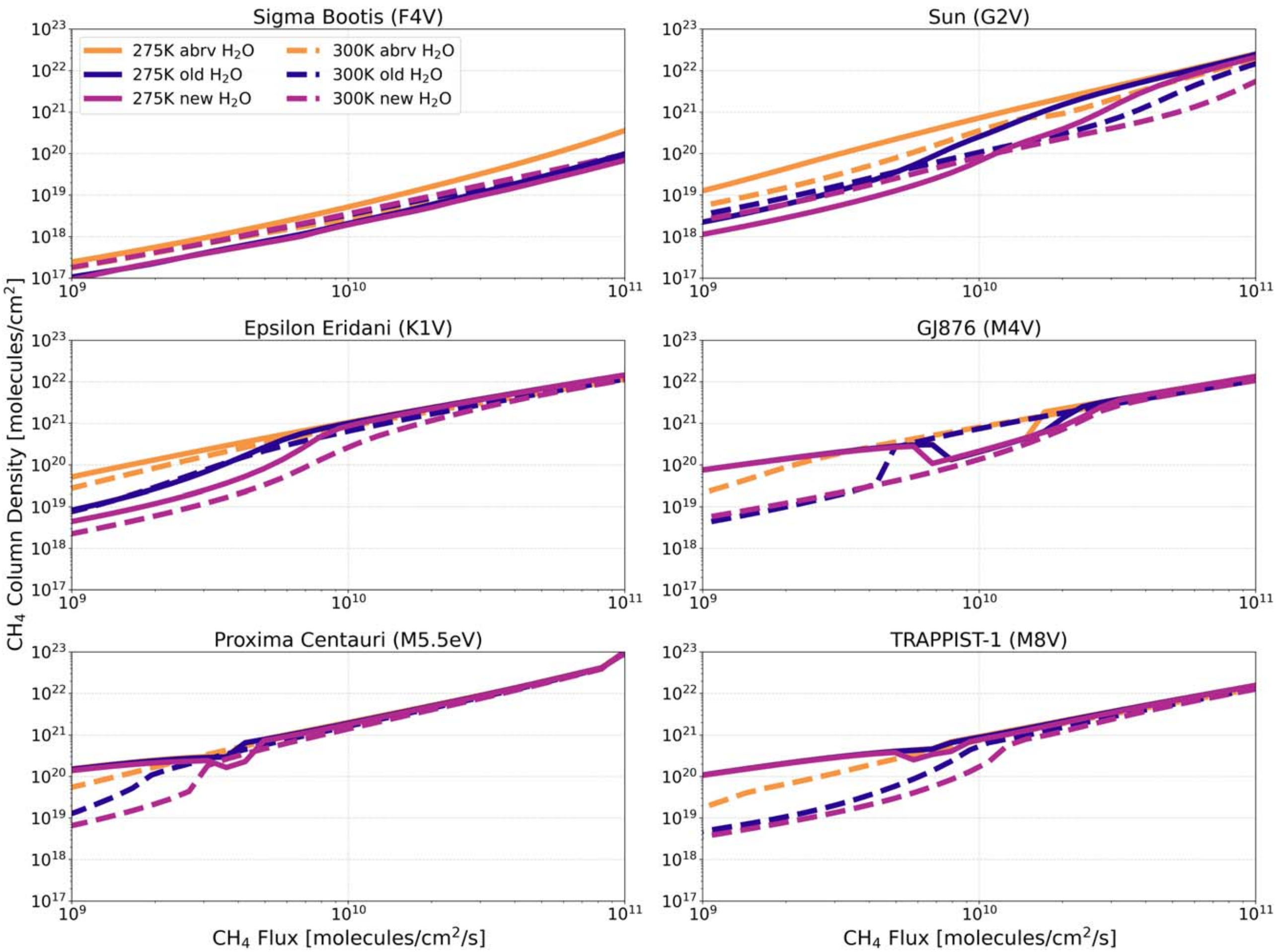

**Figure 18.** Biotic $H_2O$ cross-section sensitivity test; $CH_4$ column density vs. $CH_4$ flux for anoxic habitable planets orbiting FGKM-type host stars.

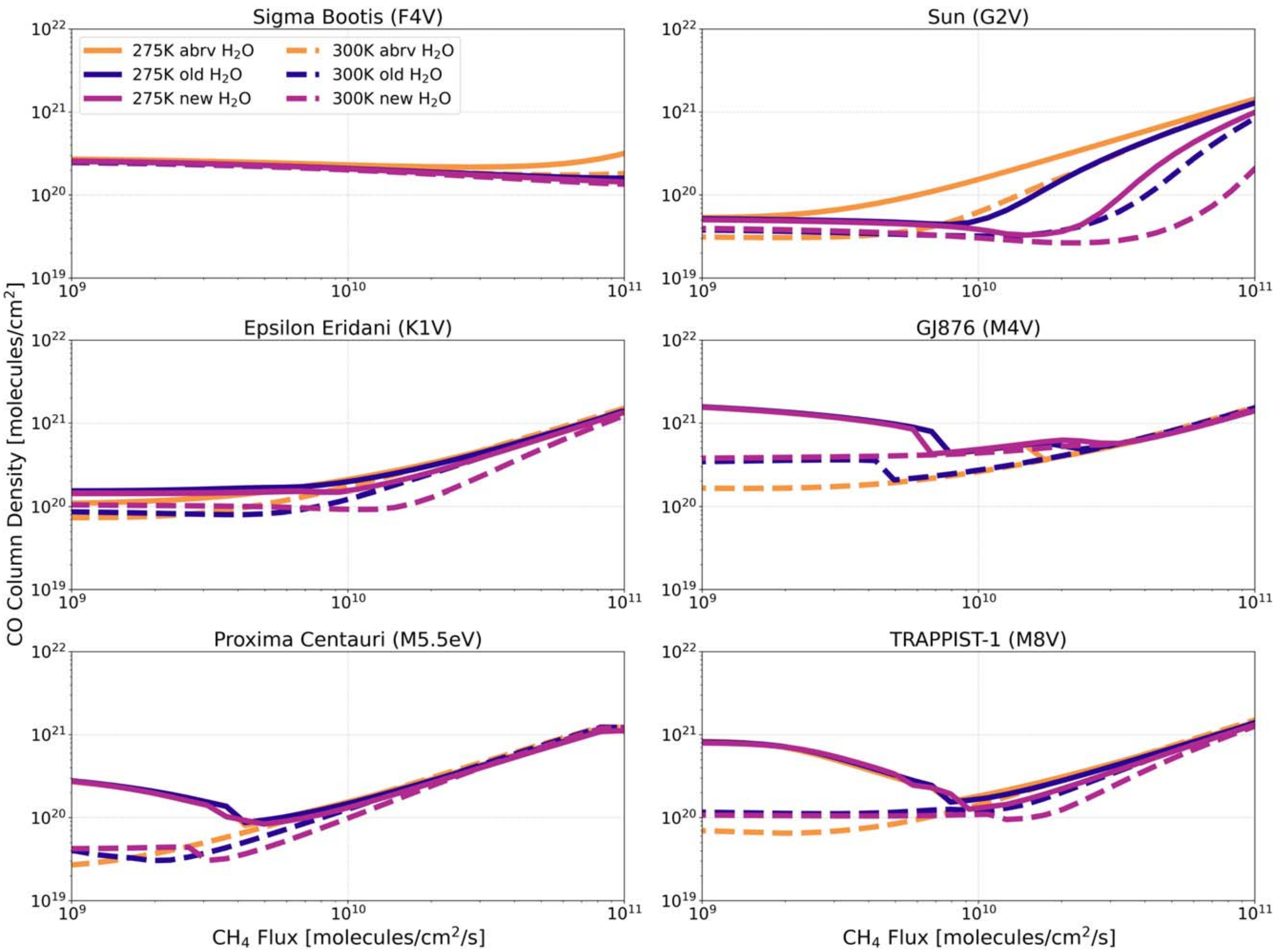

**Figure 19.** Biotic $H_2O$ cross-section sensitivity test; CO column density vs. $CH_4$ flux for anoxic habitable planets orbiting FGKM-type host stars.

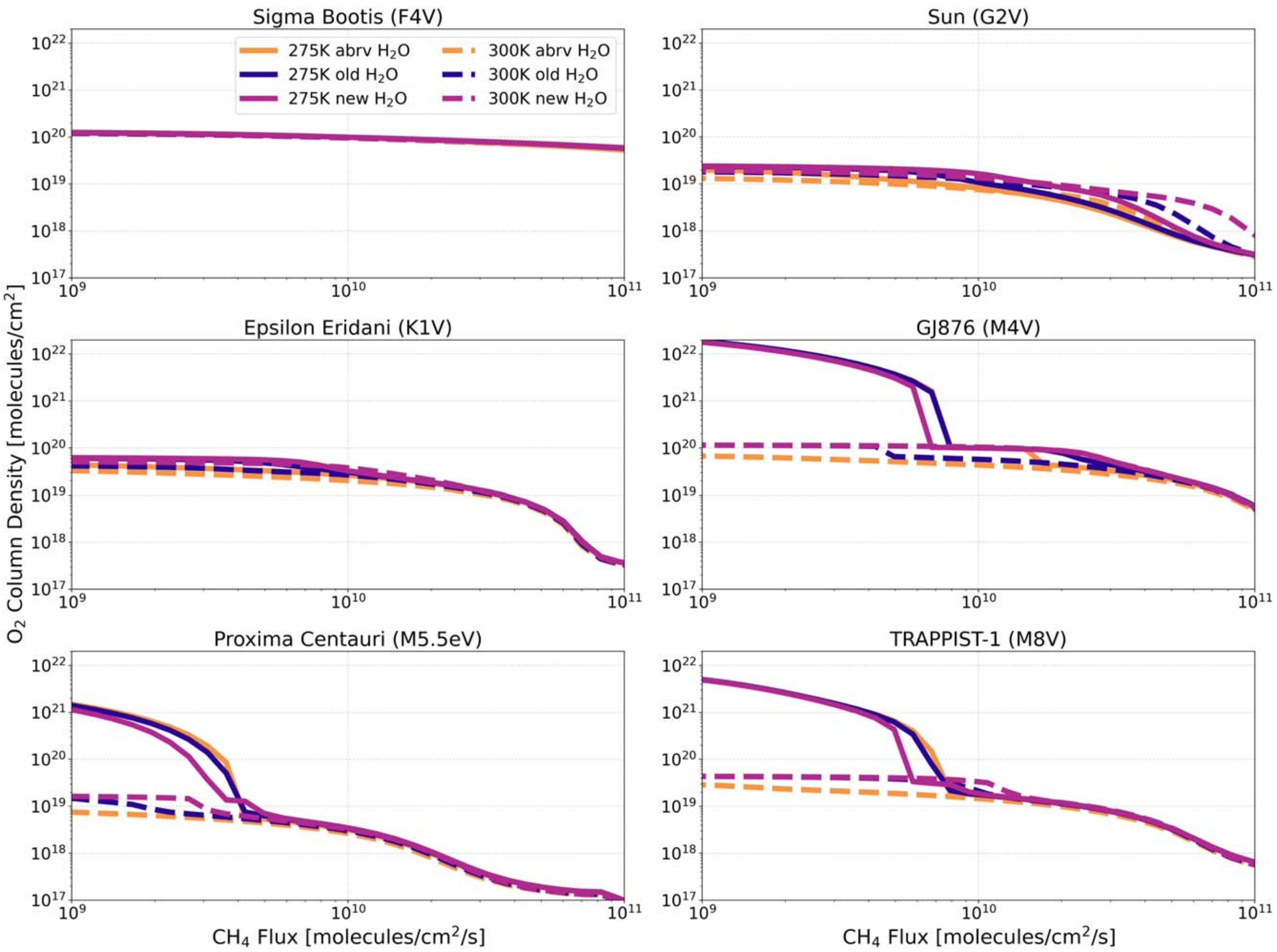

**Figure 20.** Biotic $H_2O$ cross-section sensitivity test; $O_2$ column density vs. $CH_4$ flux for anoxic habitable planets orbiting FGKM-type host stars.

## ORCID iDs

Wynter Broussard https://orcid.org/0009-0006-2304-3419
Edward W. Schwieterman https://orcid.org/0000-0002-2949-2163
Sukrit Ranjan https://orcid.org/0000-0002-5147-9053
Clara Sousa-Silva https://orcid.org/0000-0002-7853-6871
Alexander Fateev https://orcid.org/0000-0003-2863-2707